\documentclass[11pt]{article}
\usepackage{amsmath}
\usepackage{amssymb}
\usepackage{latexsym}
\usepackage{epsfig}

\newcommand{\be}{\begin{equation}}
\newcommand{\ee}{\end{equation}}

\newcommand{\Ci}{C}

\newcommand{\sectiono}[1]{\section{#1}\setcounter{equation}{0}}

\newcommand{\ra}{\rangle}
\newcommand{\la}{\langle}
\newcommand{\p}{\partial}

\begin{document}

{}~
\hfill\vbox{\hbox{hep-th/0506077}\hbox{MIT-CTP-3588}
}\break

\vskip 3.0cm

\centerline{\Large \bf A Closed String Tachyon Vacuum ?}
%

\vspace*{10.0ex}

\centerline{\large Haitang Yang
and Barton Zwiebach}

\vspace*{7.0ex}

\vspace*{4.0ex}

\centerline{\large \it  Center for Theoretical Physics}

\centerline{\large \it
Massachusetts Institute of Technology}

\centerline{\large \it Cambridge,
MA 02139, USA}
\vspace*{1.0ex}

\centerline{hyanga@mit.edu, zwiebach@lns.mit.edu}

\vspace*{10.0ex}

\centerline{\bf Abstract}
\bigskip
\smallskip

In bosonic closed string
field theory the ``tachyon potential" is a potential
for the tachyon, the dilaton, and an infinite set of
massive fields.
Earlier computations of the potential did not include the dilaton
and the  critical point formed by the quadratic and cubic interactions
was destroyed by the quartic tachyon term.
We include the dilaton contributions to the potential
 and find that a
critical point survives and appears to become more shallow.
We are led to consider the
existence of a closed string tachyon vacuum, a critical point
with zero action that represents
a state where space-time ceases to be dynamical.
Some evidence for this interpretation is found
from the study of the coupled metric-dilaton-tachyon
effective field equations, which exhibit
rolling solutions in which the dilaton
runs to strong coupling and the Einstein metric undergoes
collapse.

\vfill \eject

\baselineskip=16pt

\vspace*{10.0ex}

\tableofcontents

\sectiono{Introduction and summary}

In the last few years the instabilities associated with open
string tachyons have been studied extensively and have become
reasonably well understood~\cite{reviews}. The instabilities
associated with closed string tachyons have proven to be harder to
understand.  For the case of localized closed string tachyons --
tachyons that live on subspaces of spacetime -- there are now
plausible conjectures for the associated instabilities and a fair
amount of circumstantial evidence for
them~\cite{localized,Okawa:2004rh,Bergman:2004st,Adams:2005rb,
Suyama:2005hw}.

The bulk tachyon of the closed bosonic string is the
oldest known closed string tachyon.  It remains
 the most mysterious one and
 there is no convincing analysis of the
associated instability.
The analogy with open strings, however, suggests a fairly
dramatic possibility.  In open bosonic string in the
background of a spacefilling D-brane, the tachyon potential
has a critical point that represents spacetime without
the D-brane and thus without physical
open string excitations.  In an analogous closed string
tachyon vacuum one would expect no closed string excitations.
Without gravity excitations spacetime ceases to be dynamical
and it would seem that, for all intents and purposes,
it has dissappeared.

There has been no consensus that such a closed string
tachyon vacuum exists.  In fact, no analysis of the closed
string tachyon potential (either in the CFT approach or in the SFT approach)
has provided concrete evidence of a vacuum with non-dynamical spacetime.
Since the analogous open string tachyon vacuum shows
up quite clearly in the open string field theory computation
of the potential it is natural to consider the corresponding
calculation in closed string field
theory (CSFT)~\cite{Zwiebach:1992ie,Saadi:tb}.

The quadratic and cubic terms
in the closed string tachyon potential are well
known~\cite{Kostelecky:1990mi,Belopolsky:1994sk}:
\begin{equation}
\label{cubpot}
\kappa^2\mathbb{V}^{(3)}_0 = - t^2  +
\frac{6561}{4096}\,t^3   \,,    \quad (\alpha' = 2)\,.
\end{equation}
These terms define a critical point  analogous to
the one that turns out to represent the tachyon vacuum in the open
string field theory. In open string field theory higher
level computations make the vacuum about 46\% deeper. Since CSFT
is nonpolynomial, it is natural to investigate the effect of
the quartic term in the potential. This term was
found to be~\cite{Belopolsky:1994bj,Moeller:2004yy}
\begin{equation}
\label{v4efft}
\kappa^2V^{(4)}_0 =-3.0172\, t^4 \,.
\end{equation}
This term is so large and negative that $\mathbb{V}^{(3)}_0+ V_0^{(4)}$ has no
critical point. In fact, the quartic term in the {\em effective}
tachyon potential (obtained by integrating out massive fields)
is even a bit larger~\cite{Belopolsky:1994bj}.
The hopes of identifying a reliable critical
point in the closed string tachyon potential were
dashed\footnote{In the effective open string tachyon potential
a negative quartic term
also destroys the cubic critical point.
Nevertheless,
the critical point can be gleaned using Pade-approximants~\cite{Taylor:2002fy}.
For closed strings, however, the quartic term is too large: for a potential
 $v(t) = v_2 t^2 + v_3 t^3 + v_4 t^4$, with $v_2, v_4<0$, the approximant
formed by the ratio of a cubic and a linear
polynomial fails to give a critical point when $v_2 v_4 \geq v_3^2$.}.

Recent developments inform our present analysis. The tachyon potential
must include all fields that are sourced by the zero-momentum
tachyon. As discussed in~\cite{Sen:1999xm}, this includes massless
closed string states that are built from ghost oscillators, in particular,
the zero-momentum
ghost-dilaton state $(c_1 c_{-1} - \bar{c}_1 \bar{c}_{-1})|0\rangle$.
The search for a critical point
cannot be carried out consistently
without including the ghost dilaton.  Computations
of quartic vertices coupling dilatons, tachyons, and other massive
fields are now possible due to the work of Moeller~\cite{Moeller:2004yy}
and have been done to test the marginality of matter and dilaton
operators~\cite{Yang:2005iu,Yang:2005ep}.

As we explain now, ghost-dilaton couplings to the tachyon restore
the critical point in the potential.
The key  effect can be understood from the
cubic and quartic couplings
\begin{equation}
\kappa^2 V (t, d) = - {27\over 32}\, \,t\, d^2\,+\,3.8721 \,t^3 d  + \ldots \,\,.
\end{equation}
The cubic coupling  plays no role as long as we only consider cubic
interactions: $d$ can be set consistently to zero. The quartic
coupling is linear in $d$.  Once included,
the equation of motion for the dilaton can
only be satisfied if the dilaton acquires an expectation value.
Solving for the dilaton one finds $d = 2.2944\, t^2$ and
substituting back,
\begin{equation}
\kappa^2V (t, d) = 4.4422\, t^5 + \ldots
\end{equation}
This positive quintic term suffices to compensate the effects
of (\ref{v4efft}) and restores the critical point.  Our computations
include additional couplings and the effect of massive fields as well.
The critical point persists and may be reliable, although more work is
needed to establish this convincingly.

In order to interpret the critical point
 we raise and answer a pair of questions.
The ghost-dilaton has a
positive expectation value at the critical point.
Does this correspond to stronger or weaker
string coupling ?  We do a detailed comparison of quadratic and cubic
terms in the closed string field theory action and in the low-energy
effective field
theory action.  The conclusion is that the positive
dilaton expectation value corresponds
to {\em stronger} coupling. In our solution the ghost-dilaton is excited
but the scalar operator $c\bar c\, \partial X \cdot \bar\partial X $, sometimes
included in the dilaton vertex operator, is not. We ask: Is the
string metric excited? Is the Einstein metric excited?  These questions
are only well-defined at the linearized level, but the answers are clear:
the string metric does {\em not} change, but the Einstein metric
does.  We take the opportunity to  explain the relations between
the four kinds of ``dilatons" that are used in the literature: the
ghost-dilaton,
the matter-dilaton, the dilaton, and the dilaton of the older literature.
It is noted that one cannot define unambiguously
a dilaton vertex operator unless one specifies which
 metric is left invariant; conversely,
the metric vertex operator is only determined once one specifies which dilaton
is left invariant.

In a companion paper~\cite{HZR} we attempted to gain insight
into the tachyon vacuum  by considering the
rolling solutions\footnote{Rolling solutions have long been considered using
Liouville field theory to provide conformal invariant
sigma model with spacetime background
fields that typically include a linear dilaton and a constant string
metric~\cite{Tseytlin:1990mz,Strominger:2003fn,Kluson:2003xn,DaCunha:2003fm}.
} of a low-energy effective action  for
the string metric $g_{\mu\nu}$, the tachyon $T$, and the dilaton
$\Phi$:
 \begin{equation}
\label{sigma_action}
S_\sigma=\frac{1}{2\kappa^2 }\int \, d^D x
\sqrt{- g}\, e^{-2\Phi}\Bigl(R+4 (\p_\mu\Phi)^2  -
(\p_\mu T)^2  -2 V(T)\,\Bigr)\,.
\end{equation}
This action, suggested by the beta functions of
sigma models with background fields~\cite{sdas},
 is expected to capture at least
some of the features of string theory solutions. 
The potential
is tachyonic: $V(T) = -{1\over 2} m^2 T^2 + \mathcal{O}(T^3)$,
but is otherwise left undetermined.  We found that solutions in
which the tachyon begins the rolling process always have
{\em constant} string metric for all times -- consistent with the type of
the SFT critical point.  The dilaton, moreover, grows
in time throughout the evolution -- consistent with the larger
dilaton vev in the SFT critical point.
Rather generally, the solution becomes
singular in finite time: the dilaton runs to infinity and the string
coupling becomes infinite. Alternatively, the Einstein metric crunches
up and familiar spacetime  no longer exists. This
seems roughly consistent with the idea that the tachyon vacuum
does not have a fluctuating spacetime.

Perhaps the most subtle point concerns the value
of the on-shell action. In the open string field theory computation
of the tachyon potential, the value of the action (per unit spacetime
volume) is energy density.  The tachyon
conjectures are in fact formulated in terms of energy
densities at the perturbative and the non-perturbative vacuum~\cite{reviews}.
Since the tree-level cosmological constant in closed
string theory is zero, the value of the action at the perturbative
closed string vacuum is zero.  We ask: What is the value of the
potential, or action (per unit volume) at the critical point ?
The low-energy action (\ref{sigma_action}) suggests a surprising
answer. Consider the associated equations of motion:
\begin{equation}
\label{smeqmotion}
\begin{split}
R_{\mu\nu} +
2 \nabla_\mu\nabla_\nu \Phi  - (\p_\mu T) (\p_\nu T) &= 0\,, \\[0.5ex]
\nabla^2 T - 2 (\p_\mu \Phi)  (\p^\mu T) - V' (T) &= 0\,, \\[0.5ex]
\nabla^2 \Phi - 2 (\p_\mu \Phi)^2 - V(T) &= 0\,.
\end{split}
\end{equation}
If the fields acquire {\em constant} expectation values we can satisfy
the tachyon equation if the
expectation value $T_*$ is a critical point of the potential:
$V'(T_*) =0$. The dilaton equation imposes an
additional constraint: $V(T_*)=0$, the potential must itself
vanish. This is a reliable constraint that follows from a simple fact:
in the action the dilaton appears without derivatives only
as a multiplicative factor. This fact remains true after
addition of $\alpha'$ corrections of all orders.  It may
be that $V(T)$ has a critical point $T_0$ with $V(T_0) <0$,
but this cannot be the tachyon vacuum.  The effective field
equations imply that a vacuum with spacetime
independent expectation values has {\em zero} action.

The
action (\ref{sigma_action}) can be evaluated on-shell
using the equations of motion. One finds
 \begin{equation}
\label{sigma_action_os}
S_{on-shell}=\frac{1}{2\kappa^2 }\int \, d^{\,d+1} x
\sqrt{- g}\, e^{-2\Phi}\bigl( -4 V(T)\,\bigr)\,.
\end{equation}
In  rolling solutions the action density changes
in time but,  as $\Phi\to\infty$
at late times the action density
goes to zero~\cite{HZR}. This also
suggests that the tachyon vacuum is a critical point with zero
action.

\begin{figure}[!ht]
\leavevmode
\begin{center}
\epsfysize=5.8cm
\epsfbox{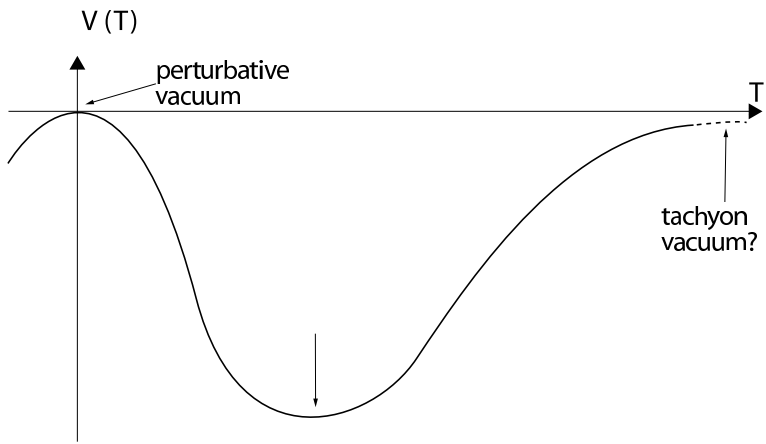}
\end{center}
\caption{\small A sketch of a closed string tachyon potential consistent
with  present evidence. The perturbative
vacuum is at  $T=0$. The closed string tachyon vacuum would be the critical
point with zero cosmological term, shown here at $T\to \infty$ (in CSFT
this point corresponds to finite tachyon vev).
A critical point with negative cosmological constant cannot
provide a spacetime independent tachyon vacuum.}
\label{tpot_conj}
\end{figure}

In Figure~\ref{tpot_conj} we present the likely  features of the
tachyon potential.  The unstable perturbative vacuum $T=0$
has zero cosmological constant, and so does the tachyon vacuum
$T=\infty$.  The infinite value of $T$ is suggested
by the analogous result in the effective open string theory
tachyon potential (see conclusions).  In SFT the tachyon vacuum
 appears for finite values of the fields, but
the qualitative features would persist.
The potential is qualitatively in the class used
 in cyclic universe models~\cite{Steinhardt:2004gk}.

In our calculations we  find some evidence that the action
density, which is negative,  may go to zero as we increase
the accuracy of the calculation.  To begin with, the value $\Lambda_0$
of the action density at the critical point of the cubic
tachyon potential (\ref{cubpot})
may be argued to be rather small.  It is a cosmological term about
seventy times smaller than the ``canonical" one associated with
$D=2$ non-critical string theory (see \cite{Okawa:2004rh}, footnote 5).
Alternatively, $\Lambda_0$ is only about 4\% of the value that would
be obtained using the on-shell coupling of three tachyons
to calculate the cubic term.  The inclusion of
cubic interactions of massive fields makes the action density
about 10\% more negative.  This shift, smaller than the
corresponding one in open string field theory,
is reversed once we include the dilaton quartic terms.  In the most
accurate computation we have done, the action density is down to
60\% of~$\Lambda_0$.  Additional computations are clearly in order.

\smallskip
As a by-product of our work, we  investigate
large dilaton deformations in CSFT.  For
ordinary marginal deformations the description reaches
an obstruction for some finite critical value of the string field marginal
parameter~\cite{Sen:2000hx,Sen:2004cq}. The critical value is stable
under level expansion, and
the potential for the marginal field (which should vanish for infinite level)
is
small.
For the dilaton, however, the lowest-order obstruction
is not present~\cite{Yang:2005ep}. We  carry this analysis to higher
order and no reliable obstructions are found:
critical values of the dilaton jump wildly with
level and appear where the dilaton potential
is large and cannot be trusted.  This result strengthens
the evidence that CSFT can describe backgrounds with arbitrarily large
variations
in the string coupling.  If the infinite string coupling limit
is also contained in the configuration space it may be possible
to define M-theory using type IIA superstring field theory.

\medskip
Let us briefly describe the contents of this paper.
In section 2 we
reconsider the universality arguments~\cite{Sen:1999xm} that require the
inclusion of the ghost-dilaton, exhibit a world-sheet parity symmetry
that allows a sizable truncation of the universal space, 
and note that universality
may apply in circumstances significantly more general that originally
envisioned~\cite{SenZwiebach}.  Our computational strategy
for the tachyon potential,
motivated by the results of~\cite{Yang:2005iu,Yang:2005ep},
goes as follows.  We compute {\em all}
quadratic and cubic terms in the
potential including fields up to level four.  We then begin
the inclusion of quartic terms and obtain complete results up
to quartic interactions of total level four.
The results make it plausible that a critical point exists and that the
value of the action density  decreases in magnitude as the
accuracy improves.   In section 3 we find the linearized
relations between the metric, dilaton, and tachyon closed string fields
and the corresponding fields in the sigma-model approach to
string theory.  These relations allow us to establish that
the dilaton vev at the critical point represents an increased string
coupling and that the string field at the critical point  does not
have a component along the vertex operator for the
string metric.  We discuss the vertex operators
associated with the various definitions of the dilaton,
determine the nonlinear field relations between
the string field theory and effective field theory dilatons and tachyons
to quadratic order and at zero-momentum, and examine large dilaton
deformations.
In the concluding section we discuss additional considerations
that suggest the existence of the tachyon vacuum. These come from
non-critical string theory, p-adic strings, and sigma model
arguments.  Finally, the
details of the nontrivial computations of quartic couplings
are given in the Appendix.

\sectiono{Computation of the tachyon potential}

In this section we present the main computations of this paper.
We begin by introducing the  string field relevant for the
calculation of the tachyon potential, giving
a detailed discussion of universality. This string field contains
the tachyon, at level zero, the ghost-dilaton, at level two, and
massive fields at higher even levels.   We then give  the quadratic
and cubic couplings for the string field restricted to level four
and calculate the critical point.
Finally, we give the quartic couplings at level zero, two, and
four.  The critical point survives the inclusion of quartic
interactions and becomes more shallow -- consistent with the
conjecture that the tachyon vacuum has zero action.

The computations use the closed string field
action~\cite{Zwiebach:1992ie,Saadi:tb,Okawa:2004rh}, which takes the form
\begin{equation}
\label{csft_action}
S=-\frac{2}{\alpha'}\Big(\frac{1}{2} \langle \Psi|c_0^-\,
Q|\Psi\rangle +\frac{\kappa}{3!} \{ \Psi,
\Psi\,,\Psi\} +\frac{\kappa^2}{4!} \{
\Psi,\,\Psi,\Psi,\Psi\} +\cdots\Big).
\end{equation}
The string field $\Psi$
lives on  $\mathcal{H}$,
the ghost number two state space
of the full CFT restricted to the
subspace of states that satisfy
\begin{equation}
\label{subcond}
(L_0 - \bar L_0) |\Psi\rangle = 0 \, \quad \hbox{and}\quad
(b_0 - \bar b_0) |\Psi\rangle = 0 \,.
\end{equation}
The BRST operator is $Q= c_0L_0 + \bar c_0 \bar L_0 + \dots$, where
the dots denote terms independent of $c_0$ and of~$\bar c_0$.
Moreover, $c_0^\pm = {1\over 2} (c_0 \pm \bar c_0)$,
and we normalize correlators using $ \langle 0| c_{-1}\bar c_{-1} c_0^- c_0^+ c_1
\bar c_1|0\rangle=1$. All spacetime coordinates are imagined
compactified with the
volume of spacetime set equal to one.

\subsection{Tachyon potential universality and the ghost-dilaton}

The universality of the closed string tachyon potential was
briefly discussed in~\cite{Sen:1999xm}, where it was also
noted that the ghost number
two universal string field that contains the tachyon
should include the zero-momentum ghost-dilaton
state $(c_{1} c_{-1} - \bar c_1 \bar c_{-1} )  |0\rangle$. In here
we review the universality argument and extend it slightly, offering
the following  observations:

\begin{itemize}

\item  The ghost-dilaton must be included because closed string
field theory is not cubic.

\item A  world-sheet parity symmetry of closed
string field theory can be used to restrict
the universal subspace.

\item  The arguments of~\cite{Sen:1999xm} do not apply directly
to general CFT's, linear dilaton backgrounds, for example.
If the closed string background is defined by a general matter
CFT, solutions on the universal subspace may still be solutions,
but there is no tachyon potential~\cite{SenZwiebach}.

\end{itemize}

The original idea in universality is to produce a subdivision
of all the component fields of the string field theory
into two disjoint sets, a set $\{ t_i\}$
that contains the zero-momentum tachyon
and a set $\{u_a\}$ such that the string field action
$S(t_i, u_a)$ contains no term with a single $u$-type field.
It is then consistent to search
for a solution of the equations of motion that assumes $u_a=0$
for all $a$.

To produce the desired set $\{t_i\}$ we assume that the matter
CFT is such that $X^0$ is the usual negative-metric
field with associated conserved momentum $k_0$ and the rest
of the matter CFT is unitary.
The state space $\mathcal{H}$ (see (\ref{subcond})) is then  divided
into three disjoint vector subspaces
$\mathcal{H}_1,  \mathcal{H}_2,$ and  $\mathcal{H}_3$. One has
$\mathcal{H}_i = \mathcal{M}_i \otimes |\mathcal{G}\rangle$,
where $|\mathcal{G}\rangle$ denotes a state built with ghost
and antighost oscillators only and
$\mathcal{M}_1,  \mathcal{M}_2,$ and  $\mathcal{M}_3$ are disjoint subspaces
of the matter CFT whose union gives the total matter CFT
state space:
\begin{eqnarray}
\label{sthesplitvematt}
\mathcal{M}_1 : &&  \hbox{the~} SL(2,C) \hbox{~vacuum} ~
|0\rangle ~\hbox{and descendents} ,\nonumber\\
\mathcal{M}_2 : &&  \hbox{states with~} k_0 \not=0, \\
\mathcal{M}_3 : &&  \hbox{primaries with}~ k_0=0
~\hbox{but different from}~|0\rangle ~\hbox{and descendents} \nonumber \,.
\end{eqnarray}
In the above, primary and descendent refers to the matter Virasoro operators.
Note that the primaries in $\mathcal{M}_3$
have positive conformal dimension.  The BRST operator preserves the
conditions (\ref{subcond}), and since it
is composed of ghost oscillators
and matter Virasoro operators, it maps each $\mathcal{H}_i$
into itself. Finally, the spaces $\mathcal{H}_i$ are orthogonal
under the BPZ inner product; they only couple to themselves.

The claim is that the set $\{t_i\}$
is in fact $\mathcal{H}_1$, the states built upon the zero momentum vacuum.
The ``tachyon potential" is the string action evaluated for $\mathcal{H}_1$.

We first note
that because of momentum conservation
fields in $\mathcal{H}_2$ cannot couple linearly to
 fields  in ${\mathcal H}_1$.
 The
fields in $\mathcal{H}_3$ cannot couple linearly to the fields in
$\mathcal{H}_1$ either.   They cannot do so through the kinetic term because
the BRST operator preserves the space and $\mathcal{H}_1$ and $\mathcal{H}_3$
are BPZ orthogonal.
We also note that the matter correlator in
the $n$-string vertex does not couple $n-1$ vacua $|0\rangle$ from
$\mathcal{H}_1$ to a matter primary from $\mathcal{H}_3$:  this
is just the one-point
function of the primary in $\mathcal{H}_3$, which vanishes because the state
has non-zero dimension. The (matter) Virasoro conservation laws on the vertex
then
imply that  the coupling of any $(n-1)$ states in $\mathcal{H}_1$ to a state in
$\mathcal{H}_3$ must vanish. This completes the proof that $\mathcal{H}_1$
is the subspace for tachyon condensation.

The space $\mathcal{H}_1$ can be written
as
\begin{equation}
\label{biganswer}
\hbox{Span} \Bigl\{ L_{-j_1}^m \ldots L_{-j_p}^m\,
\bar L_{-\bar j_1}^m \ldots \bar L_{-\bar j_{\bar p}}^m
b_{-k_1} \ldots b_{-k_q}\, \bar b_{-\bar k_1} \ldots \bar b_{-\bar k_{\bar q}}
\, c_{-l_1} \ldots c_{-l_r} \,
\, \bar c_{-\bar l_1} \ldots \bar c_{-\bar l_{\bar r}} \, |0\rangle\Bigr\}\,,
\end{equation}
where
\begin{equation}
j_1 \geq j_2 \geq \ldots \geq j_p \,, ~~j_i \geq 2 \,, ~~
\bar j_1 \geq \bar j_2 \geq \ldots \geq \bar j_{\bar p }\,, ~~\bar j_i \geq
2\,,
\end{equation}
as well as
\begin{equation}
 k_i, \bar k_i \geq  2
\,, ~~ l_i, \bar l_i \geq
  -1\,,
\quad
\hbox{and}\quad  r+ \bar r - q - \bar q = 2 \,.
\end{equation}
Finally, the states above must also be annihilated by $L_0- \bar
L_0$ as well as $b_0 - \bar b_0$.

There is a reality condition on
the string field~\cite{Zwiebach:1992ie}: its BPZ and hermitian conjugates must
differ by a sign.  We show now
that this condition
is satisfied by all the states in (\ref{biganswer}), so
the coefficients by which they are multiplied in the
universal string field (the zero-momentum spacetime fields)
must be real.
Suppose a state is built with $p$
ghost oscillators and $p-2$ antighost oscillators. The BPZ
and hermitian conjugates differ by the product of two
factors: a  $(-1)^p$ from
the BPZ conjugation of the ghost oscillators and a
$(-1)^{(2 p-2)(2p-1)/2}= (-1)^{p-1}$ from the reordering of oscillators
in the hermitian conjugate.  The product of these two factors
is minus one, as we wanted to show.

\medskip
In open string theory twist symmetry, which arises from
world-sheet parity, can be used to further
restrict the universal subspace constructed from matter Virasoro
and ghost oscillators.   In the case of closed string theory
the world-sheet parity transformation that exchanges holomorphic
and antiholomorphic sectors is the relevant symmetry.\footnote{We thank
A.~ Sen for discussions that led us to construct the arguments
presented below.} World-sheet
parity is not necessarily a symmetry of arbitrary matter CFT's, but
it is a symmetry in the universal subspace:
correlators are complex
conjugated when we exchange holomorphic and antiholomorphic
Virasoro operators
as $T(z) \leftrightarrow \bar T(\bar z)$.
More precisely, we introduce
a $\star$-conjugation, a map of $\mathcal{H}_1$ to $\mathcal{H}_1$
that is an involution.
In a basis of Virasoro modes $\star$ can be written explicitly as the map
of states
\begin{equation}
\label{rule_matter}
\star \, :  \quad A\, L_{-i_1} \cdots L_{-i_n}\, \bar L_{-j_1}
\cdots \bar L_{-j_n} |0\rangle \quad \to
\quad  A^*\,\bar L_{-i_1} \cdots \bar L_{-i_n} \,L_{-j_1}
\cdots L_{-j_n} |0\rangle\,,
\end{equation}
where $A$ is a constant and $A^*$ denotes its complex conjugate.  Given
the operator/state correspondence, the above defines completely the
star operation $\star \,: \mathcal{O} \to \mathcal{O}^\star$
on vertex operators for vacuum descendents. It
 results in the following property for the correlator of $n$ such operators
placed at $n$ points on a Riemann surface:
\begin{equation}
\langle \mathcal{O}_1 \ldots  \mathcal{O}_n\rangle =
\langle \mathcal{O}_1^\star \ldots  \mathcal{O}_n^\star\rangle^* \,.
\end{equation}
In the ghost sector of the CFT a small complication with
signs arises because the basic correlator is odd under the
exchange of holomorphic and anti-holomorphic sectors:
\begin{equation}
\label{faxsorr}
\langle\,  c(z_1)c(z_2)c(z_3) \,   \bar c (\bar w_1)\bar c (\bar
w_2)\bar c (\bar w_3)
\rangle = -\langle \, \bar c(\bar z_1)\bar c(\bar z_2) \bar c(\bar z_3) \,
 c (w_1) c (w_2) c ( w_3)\,
\rangle^* \,.
\end{equation}
Since two-point functions
of the ghost fields are complex conjugated by the exchanges
$c(z) \leftrightarrow \bar c (\bar z)$
and $b(z) \leftrightarrow \bar b (\bar z)$, it
follows from (\ref{faxsorr}) that performing these exchanges
on an {\em arbitrary} correlator of ghost and antighost fields will give minus the
complex conjugate of the original correlator.
We will define  $\star$-conjugation in the ghost sector by:
\begin{equation}
\label{rule_ghost}
\star \, :  \quad A\, c_{i_1}\hskip-2pt\cdot\cdot c_{i_n}\,
b_{j_1} \hskip-2pt\cdot\cdot b_{j_m}\, \bar c_{k_1}
\hskip-2pt\cdot\cdot {\bar c}_{k_r}\,
{\bar b}_{l_1}\hskip-2pt\cdot\cdot {\bar b}_{l_s} |0\rangle
\,\quad \to\quad
A^*\, {\bar c}_{i_1}\hskip-2pt\cdot\cdot {\bar c}_{i_n}\,
\bar b_{j_1} \hskip-2pt\cdot\cdot \bar b_{j_m}\, c_{k_1}
\hskip-2pt\cdot\cdot c_{k_r}\,
b_{l_1}\hskip-2pt\cdot\cdot b_{l_s} |0\rangle \,.
\end{equation}
For a general state $\Psi$ of the universal subspace
we define $\Psi^\star$ to be the state obtained by
the simultaneous application of
(\ref{rule_matter}) and (\ref{rule_ghost}).
It is clear from the above discussion that the correlators satisfy
\begin{equation}
\label{n_point}
\langle \Psi_1 \, \Psi_2 \ldots \Psi_n \rangle = -
\langle \Psi_1^\star \, \Psi_2^\star \ldots \Psi_n^\star \rangle^*  \,, \qquad
\Psi_i \in \mathcal{H}_1 \,.
\end{equation}

We  now  define the action of the world-sheet parity operation $\mathcal{P}$
on arbitrary states of the universal subspace:
\begin{equation}
\label{parity_def}
\mathcal{P} \Psi\equiv - \Psi^\star, \quad  \Psi \in \mathcal{H}_1\,.
\end{equation}
We claim that the
string field theory action, restricted to $\mathcal{H}_1$,
is  $\mathcal{P}$ invariant:
\begin{equation}
\label{lkkekjj}
S (\Psi) = S (\mathcal{P} \Psi )\, , \quad \hbox{for} \quad  \Psi \in \mathcal{H}_1\,.
\end{equation}
First consider the invariance of the cubic term.  Using (\ref{parity_def})
and (\ref{n_point}) we have
\begin{equation}
\langle \mathcal{P}\Psi \,, \mathcal{P}\Psi\,,  \mathcal{P}\Psi\rangle
= - \langle \Psi^\star \,, \Psi^\star\,,  \Psi^\star\rangle
=  \langle \Psi \,, \Psi\,,  \Psi\rangle^*
=\langle \Psi \,, \Psi\,,  \Psi\rangle\,,
\end{equation}
where in the last step we used the reality of the string field action.
The kinetic term of the action is also invariant.  First note that
$ (c_0^- Q \Psi)^\star
= -c_0^- Q \Psi^\star \,.$
It then follows that
\begin{equation}
\langle \mathcal{P}\Psi\,, c_0^- Q \mathcal{P}\Psi\rangle
 = \langle \Psi^\star\,, c_0^- Q \Psi^\star\rangle =
- \langle \Psi^\star\,, (c_0^- Q \Psi)^\star\rangle=
 \langle \Psi\,, c_0^- Q \Psi\rangle^* =
\langle \Psi\,, c_0^- Q \Psi\rangle\,.
\end{equation}
For higher point interactions, the
invariance follows because the antighost insertions have the
appropriate structure.  Each time we add a new string field we must
add two antighost insertions. For the case of quartic interactions
they take the form of two factors $\mathcal{B} \mathcal{B}^\star$ (see
eqn.~(\ref{exptwoforms})). Since $(\mathcal{B} \mathcal{B}^\star)^\star
= -\mathcal{B} \mathcal{B}^\star$, the extra minus sign cancels against
the minus sign from the extra string field.  This can be seen to
generalize to higher
order interactions using the forms of the off-shell amplitudes discussed
in section 6 of~\cite{Belopolsky:1994sk}.  This completes our proof
of (\ref{lkkekjj}).

Since $\mathcal{P}^2=1$ the space $\mathcal H_1$ can be divided into two disjoint
subspaces:  the space $\mathcal H_1^+$ of states with $\mathcal{P}=1$
and the space $\mathcal H_1^-$
of states with $\mathcal{P}=-1$:
\begin{eqnarray}
\mathcal{P} (\Psi_+)&=&+\Psi_+, \hspace{7mm} \Psi_+\in \mathcal
H_1^+\nonumber\,,\\[0.5ex]
\mathcal{P} (\Psi_-)&=&- \Psi_-, \hspace{7mm} \Psi_-\in \mathcal
H_1^-\,.
\end{eqnarray}
It follows from the invariance of the action that
 no term in the action can contain just one state in $\mathcal H_1^-$.
We can therefore restrict
ourselves to the subspace $\mathcal H_1^+$ with positive parity.

The string field is further restricted
by using a gauge fixing condition.  The
computation of the potential is done in the Siegel gauge, which
requires states to be annihilated by $b_0+\bar b_0$.  To
restrict ourselves to the Siegel gauge we take the states in
(\ref{biganswer}) that have neither a $c_0$ nor a $\bar c_0$.

The Siegel gauge fixes the gauge symmetry completely for the massive levels,
but does not quite do the job at the massless level.
There are two states with $L_0 = \bar L_0=0$ in
$\mathcal{H}_1$ that are in the Siegel gauge:
\begin{equation}
(c_{1} c_{-1} - \bar c_1 \bar c_{-1} )|0\rangle \quad \hbox{and}
\quad (c_{1} c_{-1} + \bar c_1 \bar c_{-1} )|0\rangle \,.
\end{equation}
The first state is the ghost dilaton and it is proportional to $Q
(c_0- \bar c_0)|0\rangle$. Since $(c_0- \bar c_0)|0\rangle$ is not
annihilated by $b_0-\bar b_0$ the gauge parameter is illegal and
the ghost dilaton is not trivial. The second state is proportional
to $Q (c_0+ \bar c_0)|0\rangle$, so it is thus trivial at the linearized
level.  Although trivial at the linearized level, one may wonder
if the triviality holds for large fields. Happily, we need not worry:
the state is $\mathcal{P}$
odd, so it need not be included in the calculation.  The ghost-dilaton,
because of the relative minus sign between the two terms, is $\mathcal{P}$
even and it is included.

Had the closed string field theory been cubic we could have
discarded the ghost-dilaton state and all other states
with asymmetric left and right ghost
numbers. We could
restrict $\mathcal{H}_1^+$ to fields
of ghost number $(G, \bar G) = (1,1)$.  Indeed, the cubic vertex
cannot couple two $(1,1)$ fields to anything except another $(1,1)$
field.
Moreover, in the Siegel gauge $c_0^-Q$ acts as an operator of ghost
number $(1,1)$, so again, no field with asymmetric ghost numbers
can couple linearly.
 The quartic and higher order interactions in CSFT have
antighost insertions that do not have equal left and right ghost
numbers. It follows that these higher order vertices can couple
the ghost-dilaton to $(1,1)$ fields.  Indeed, the coupling of a
dilaton to three tachyons does not vanish.  We {\em cannot} remove
from~$\mathcal{H}_1^+$ the
dilaton, nor other states with asymmetric left and right ghost
numbers.

\medskip

The  construction of the universal string field
and  action presented here does not work fully if the
matter CFT contains  a linear dilaton background.
Momentum conservation along the corresponding
coordinate is anomalous and one cannot build
an action with states of zero momentum
only: the action restricted to $\mathcal{H}_1$ is identically zero.
There would be no universal ``potential" in $\mathcal{H}_1$.
It appears rather likely, however, that any {\em solution} in
the universal subspace would still be a solution in a
linear dilaton background. In fact, any solution in
the universal subspace may be a solution
for string field theory formulated with a  general
matter CFT~\cite{SenZwiebach}.

\medskip
We conclude this section by writing out the string field for the first few levels.
The level $\ell$ of a state is defined by
$\ell =L_0+\bar L_0+2\,.$
The level zero part of the string field is
\begin{equation}
|\Psi_0\rangle =t\, c_{1}\bar c_1 |0\rangle\,.
\end{equation}
Here $t$ is the zero-momentum tachyon. The level two part of the string field
is
\begin{equation}
|\Psi_2\rangle =d\, (c_{1} c_{-1} - \bar c_1 \bar c_{-1} )  |0\rangle\,.
\end{equation}
Here $d$ is the zero momentum ghost-dilaton.  It multiplies the only state of
$\mathcal{P} = +1$ at this level.
At level four there are four component fields:
\begin{eqnarray}
|\Psi_4\rangle &=&\Bigl( f_1\, c_{-1}\bar c_{-1}+\,f_2\, L_{-2}
c_1 \,\bar L_{-2}\bar c_1 +\,f_3\, (L_{-2} c_1\bar c_{-1}+
          c_{-1}\,\bar L_{-2}\bar c_1) \nonumber \\[0.5ex]
&& \,+\,g_1\,(b_{-2} c_1\, \bar c_{-2}\bar c_1 \,-\, c_{-2}c_1\,
\bar  b_{-2} \bar c_1)\Bigr) |0\rangle\,.
\end{eqnarray}
Note that the states coupling to the component fields all have
 $\mathcal{P}=+1$ and that $g_1$ couples to a state with
asymmetric left and
right ghost numbers. In this paper we will not use higher level
terms in the string field.

With $\alpha'=2$ the closed string field potential $V$ associated
with the action in (\ref{csft_action}) is
\begin{equation}
\label{csftpot}
\kappa^2 V=\frac{1}{2}\langle \Psi|c_0^-\, Q|\Psi\rangle
+\frac{1}{3!} \{ \Psi,\Psi,\Psi\} +\frac{1}{4!} \{
\Psi,\Psi,\Psi,\Psi\} +\cdots\,\,.
\end{equation}
Here  $|\Psi\rangle = |\Psi_0\rangle+ |\Psi_2\rangle+ |\Psi_4\rangle+ \ldots$.
Our computations will not include quintic and higher order interactions in the
string action.

\subsection{The quadratic and cubic terms in the potential}

Let us now consider the potential including only the kinetic and cubic
terms in (\ref{csftpot}).  To level zero:
\begin{equation}
\kappa^2 V^{(2)}_0 = - t^2 \,, \qquad
\kappa^2 V^{(3)}_0 =  \frac{6561}{4096}\,t^3    \,.
\end{equation}
All potentials introduced in  this subsection have a superscript
that gives the order of the interaction (two for quadratic, three
for cubic, and so on), and a subscript that gives the level
(defined by the sum of levels of fields in the interaction).
  The next terms arise at level four, where we have couplings of the
tachyon to the square of the dilaton and couplings of the level four
fields to the tachyon squared:
\begin{equation}
\label{fglekf4}
\kappa^2\,V^{(3)}_4 =  -
\frac{27}{32}\,d^2\,t + \Bigl(\, \frac{3267}{4096}\,{f_1}
  + \frac{114075}{4096} \,{f_2}   - \frac{19305}{2048}
{f_3}   \, \Bigr)\, t^2 \,.
\end{equation}
At level six we can couple a level four field, a dilaton,
and a tachyon.  Only level four fields with $G\not=\overline G$ can have
such coupling, so we find:
\begin{equation}
\label{cgdt} \kappa^2\,V^{(3)}_6 =  -  \frac{25}{8}\,  {g_1}
\,t\,d\,.
\end{equation}

At level eight there are two kinds of terms.  First, we have the  kinetic
terms for the level four fields:
\begin{equation}
\label{gknfeh}
\kappa^2\,V^{(2)}_8 = ~ {f_1}^2
  \, + 169\,{{f_2}}^2
- 26\,{{f_3}}^2   - 2\,{g_1}^2 .
\end{equation}
Second, we have the cubic interactions:
\begin{equation}
\label{cftcoudk}
\begin{split}
\kappa^2\,V^{(3)}_8 &=  -\frac{ 1 }{96}\, f_1\,d^2
  -\frac{4225}{864} \,f_2 \,d^2
+ \frac{65 }{144} \,{f_3} \,d^2\\[1.0ex] &
~~+ \frac{361}{12288}\,{{f_1}}^2\,t +
\frac{511225}{55296}\,{f_1}\,{f_2}\,t +
\frac{57047809}{110592}\,{{f_2}}^2\,t + \frac{470873}{27648}
\,{f_3}^2\,t -\frac{49}{24}\,{g_1}^2\,t
\\[1.0ex] &
~~- \frac{13585} {9216}\,{f_1}\,{f_3} \,t
-\,\frac{5400395}{27648}\,{f_2}\,  {f_3} \,t  \,.
\end{split}
\end{equation}
As we can see, these are of two types: couplings of a level four
field to two dilatons
(first line)
and couplings of two level four fields to a tachyon (second and third lines).

The terms at level 10 couple two level four fields and
a dilaton.  Because of ghost number conservation, one  of the
level four fields must have $G\not= \overline G$:
\begin{equation}
\kappa^2\,V^{(3)}_{10}=-\frac{25}{5832}\big(361\, f_1+4225\,
f_2-2470\, f_3\big)\, d\, g_1 \,.
\end{equation}
Finally, at level 12 we have the cubic couplings of three level-four fields:
\begin{equation}
\begin{split}
\kappa^2\,V^{(3)}_{12}~ &= \frac{1}{4096}
\,f_1^3+\frac{1525225}{8957952} \,f_1^2 f_2-\frac{1235}{55296}
f_1^2\,f_3+\frac{6902784889}{80621568}f_1 f_2^2 \\[1.3ex]
&~-\frac{102607505}{6718464}f_1f_2f_3
+\frac{1884233}{2239488}f_1f_3^2   \\[1.1ex]
&~ +\frac{74181603769}{26873856}
f_2^3-\frac{22628735129}{13436928} \,f_2^2f_3
+\frac{4965049817}{20155392} f_2 f_3^2 \\[1.1ex]
& ~-\frac{31167227}{3359232}
f_3^3-\frac{961}{157464}\,f_1 g_1^2
-\frac{207025}{17496}\, f_2 g_1^2 +\frac{14105}{26244}\,f_3
g_1^2 \,.
\end{split}
\end{equation}

\medskip
\subsection{Tachyon vacuum with cubic vertices only}

With cubic vertices only the dilaton expectation value is zero.
In fact, only fields with $G= \overline G = 1$ can acquire
nonvanishing expectation values.
To examine the tachyon vacuum we define a series of potentials:
\begin{equation}
\label{serpot}
\begin{split}
\mathbb{V}^{(3)}_0 \equiv & ~ V^{(2)}_0 + V^{(3)}_0 \,, \\
\mathbb{V}^{(3)}_8  \equiv & ~ \mathbb{V}^{(3)}_0 + V^{(3)}_4+ V^{(3)}_6 +
V^{(2)}_8 + V^{(3)}_8  \,, \\
\mathbb{V}^{(3)}_{12} \equiv & ~ \mathbb{V}^{(3)}_8 + V^{(3)}_{10} +
V^{(3)}_{12} \,. \\
\end{split}
\end{equation}
A few observations are in order.  In all of the above potentials
we can set $d=g_1=0$. As a consequence, $V^{(3)}_6$ and
$V^{(3)}_{10}$ do not contribute.  Since the level-two dilaton
plays no role, once we go beyond the tachyon we must include level
four fields. The kinetic terms for these fields are of level
eight, so $\mathbb{V}^{(3)}_8$ is the simplest potential beyond
level zero.  With level-four fields the next potential
is~$\mathbb{V}^{(3)}_{12}$.


The critical points obtained with the potentials $\mathbb{V}^{(3)}_0$,
$\mathbb{V}^{(3)}_8$, and $\mathbb{V}^{(3)}_{12}$ are given in
Table~\ref{w/oenergy}.  We call the value of the potential
$\kappa^2\mathbb{V}$ at
the critical point the action density.  The values of the action
density follow the pattern of open string theory.  The original cubic
critical point becomes deeper.  It does so by about 10\%, a value
significantly smaller than the corresponding one in open string field theory.

\begin{table}[ht]
\begin{center}
\renewcommand\arraystretch{1.5}
\vskip 0.1in
\begin{tabular}{|c|c|c|c|c|c|}
            \hline
            \hbox{Potential}&$t$ & $f_1$& $f_2$ &  $f_3$
&  \hbox{Action density} \\
            \hline
            $\mathbb{V}^{(3)}_0$& $0.41620$ & $--$ & $--$ & $--$ &
$-0.05774$ \\
             \hline
            $\mathbb{V}^{(3)}_8$ &$0.43678$  &$-0.06502$ &
$-0.00923$ & $-0.02611$ &  $-0.06329$ \\
            \hline
            $\mathbb{V}^{(3)}_{12}$&$0.43709$
            &$-0.06709$ & $-0.00950$ & $-0.02693$ &  $-0.06338$ \\
            \hline
\end{tabular}
\caption{Vacuum solution with cubic
vertices only}\label{w/oenergy}
\end{center}
\end{table}

\subsection{Tachyon vacuum with cubic and quartic vertices}

We can now examine the quartic terms in the potential. The associated
potentials are denoted with a superscript $(4)$ for quartic and
a subscript that gives the sum of levels of the fields that enter
the term. The quartic self-coupling of tachyons has been calculated
in~\cite{Belopolsky:1994bj,Moeller:2004yy}:
\begin{equation}
\kappa^2{V}^{(4)}_0=-3.0172\, t^4  \,.
\end{equation}
With total level two we have a coupling of three tachyons
and one dilaton.  This is calculated in Appendix A.2 and
the result is
\begin{equation}
\kappa^2{V}^{(4)}_2=\,3.8721 \,t^3 d  \,.
\end{equation}
With total level four there is the coupling of two tachyons
to two dilatons (Appendix A.2) and the coupling of three tachyons
to any of the level-four fields (Appendix A.3):
\begin{equation}
\begin{split}
\kappa^2{V}^{(4)}_4 = 1.3682\, t^2 d^2
  + t^3 \bigl(  -0.4377 \, f_1-56.262 \, f_2+ 13.024\,  f_3
+\,0.2725\, g_1\bigr)\,.
\end{split}
\end{equation}
With total level six there are three types of interactions:
a tachyon coupled to three dilatons, two tachyons coupled to a
dilaton and a level-four field, and three tachyons coupled to
a level-six field.  We have only computed the first one (Appendix A.2):
\begin{equation}
\label{td^3val}
\kappa^2{V}^{(4)}_6= -\,0.9528\, t d^3  + \ldots \,.
\end{equation}
The terms that have not been computed are indicated by the dots.
Finally, the quartic self-coupling of dilatons was computed
in~\cite{Yang:2005ep},
where it played a central role in the demonstration that the
effective  dilaton
potential has no quartic term:
\begin{equation}
\label{d^4val}
\kappa^2{V}^{(4)}_8= -\,0.1056\, d^4  + \ldots\,.
\end{equation}
We use the dots to indicate the additional level eight
interactions that should be computed.

Let us now consider the potentials that can be assembled using
the above contributions.  We use the following strategy:  we include
cubic vertices to the highest possible level and then begin
to introduce the quartic couplings level by level.  The most accurate
potential with quadratic and cubic terms that we have is
$\mathbb{V}^{(3)}_{12}$ and the tachyon vacuum it contains appears
in the last line of Table~\ref{w/oenergy}. The lowest order quartic
potential that we use is therefore:
\begin{equation}
\mathbb{V}^{(4)}_0 \equiv  ~ \mathbb{V}^{(3)}_{12} + V^{(4)}_0 \,.
\end{equation}
This potential has a familiar difficulty:  the quartic self-coupling
of the tachyon is so strong that the critical point in the potential
disappears.  As we have argued, once additional terms are included
the critical point in the potential reappears.  The higher level
potentials are defined by including progressively higher level quartic
interactions:
\begin{equation}
\label{serpot}
\begin{split}
\mathbb{V}^{(4)}_2 \equiv & ~ \mathbb{V}^{(4)}_0 + V^{(4)}_2 \,, \\[0.5ex]
\mathbb{V}^{(4)}_4 \equiv & ~ \mathbb{V}^{(4)}_2 + V^{(4)}_4  \,. 
\end{split}
\end{equation}
Since our computations of $V^{(4)}_6$ and $V^{(4)}_8$ are
incomplete, the results that follow from
$\mathbb{V}^{(4)}_6\equiv   \mathbb{V}^{(4)}_4  + V^{(4)}_6$  and
$\mathbb{V}^{(4)}_8\equiv  ~ \mathbb{V}^{(4)}_6 + V^{(4)}_8$ cannot be trusted.

We are now in a position to calculate the critical points of the
potentials $\mathbb{V}^{(4)}$. In our numerical work we input the
cubic coefficients as fractions and the quartic coefficients as
the exact decimals given above (so the $t^4$ coefficient is
treated as exactly equal to $3.0172$.) Our results are given in
Table~\ref{wqu/oenergy}.  For ease of comparison, we have included
the cubic results for $\mathbb{V}^{(3)}_{12}$ as the first line.
Furthermore, we include a line for $\mathbb{V}^{(4)}_0$ even
though there is no critical point. The next potential is
$\mathbb{V}^{(4)}_2$ which contains only the additional coupling
$t^3d$. The significant result is that the critical point
reappears and can be considered to be a (moderate) deformation of
the critical point obtained with $\mathbb{V}^{(3)}_{12}$. Indeed,
while there is a new expectation value for the dilaton (and for
$g_1$), the expectation value of the tachyon does not change
dramatically, nor do the expectation values for $f_1$, $f_2$, and
$f_3$.  The critical point becomes somewhat shallower, despite the
destabilizing effects of the tachyon quartic self-couplings.

\begin{table}[ht]
\begin{center}
{\renewcommand\arraystretch{1.6}
\vskip 0.1in
\begin{tabular}{|c|c|c|c|c|c|c|c|}
           \hline
           \hbox{Potential}&$t$ & $d$& $f_1$& $f_2$ &  $f_3$&
$g_1$ & \hbox{Action density}
            \\    \hline
           $\mathbb{V}^{(3)}_{12}$&$0.43709$ & $0$
           &$-0.06709$ & $-0.00950$ & $-0.02693$ & $-- $ &  $-0.06338$ \\
           \hline
$\mathbb{V}^{(4)}_0$  &   $--$ & $--$
           & $--$ & $--$ & $--$ &  $--$ & $-- $  \\
           \hline
           $\mathbb{V}^{(4)}_2$&$0.33783$ & $0.49243$ & $-0.08007$ &
           $-0.00619$ & $-0.02607$ & $-0.10258$ &$-0.05806$ \\
           \hline
           $\mathbb{V}^{(4)}_4$&  $0.24225$ & $0.45960$ &
$-0.04528$ & $-0.00140$ &
            $-0.01233$ & $-0.07249$ & $-0.03382$ \\
           \hline
\end{tabular}}
\caption{\small Vacuum solution with cubic
and quartic vertices. We see that the magnitude of the action
density becomes smaller as we begin to include the effects of
quartic couplings. }\label{wqu/oenergy}
\end{center}
\end{table}

At the next level, where $t^2 d^2$ and $t^3M_4$ ($M_4$ denotes a level-four
field)
 terms appear, the critical point experiences some significant change.
First of all, it becomes about 40\% more shallow; the change is large
and probably significant, given the expectation that the action density
should eventually reach zero.  The tachyon expectation  changes
considerably but the dilaton expectation value
changes little.  Due to the $t^3M_4$ terms
the expectation values of some of the level four fields change dramatically.

Glancing at Table~\ref{wqu/oenergy}, one notices that
the tachyon expectation value is becoming smaller so one
 might worry that the
critical point is approaching the perturbative vacuum.
This is, of course, a possibility. If realized, it would imply that
the critical point we have encountered is an artifact of level
expansion.  We think this is unlikely. Since the dilaton seems to be relatively
stable, a trivial critical point would have to be a dilaton deformation
of the perturbative vacuum, but such deformations have negative
tachyon expectation values (see Figure~\ref{fig marginal directions}).

\medskip
At this moment we do not have full results for higher levels. The
computation of $\mathbb{V}^{(4)}_6$ would require the evaluation
of couplings of the form $t^2dM_4$ and, in principle, couplings
$t^3M_6$ of level-six fields, which we have not even introduced in
this paper. The only additional couplings we know at present are
$t d^3$, which enters in $\mathbb{V}^{(4)}_6$ and $d^4$, which
enters in $\mathbb{V}^{(4)}_8$ (see eqns.~(\ref{td^3val}) and
(\ref{d^4val})). Despite lacking terms, we calculated the
resulting vacua to test that no wild effects take place.  The
incomplete $\mathbb{V}^{(4)}_6$ leads to $t =0.35426,~ d=0.40763$
and an action density of $-0.05553$. The incomplete
$\mathbb{V}^{(4)}_8$ leads to $t =0.36853,~ d=0.40222$ and an
action density of $-0.05836$. In these results the action density
has become more negative. Given the conjectured value of the
action, it would be encouraging if the full results at those
levels
 show an action density whose
magnitude does not become larger.

One may also wonder what happens if terms of order higher
than quartic are included in the potential.  Since the tachyon
terms in the CSFT potential alternate signs~\cite{Belopolsky:1994sk},
the quintic term is positive and will help reduce the value of
the action at the critical point. The coefficient of this coupling
will be eventually needed as computations become more accurate.
The sixtic term will have a destabilizing effect. Having survived
the destabilizing effects of the quartic term, we can hope that
those of the sixtic term will prove harmless.  If, in general,
even power terms do not have catastrophic effects, it may be
better to work always with truncations of odd power.


\sectiono{The sigma model and the string field theory pictures}

In this section we study the
relations between
the string field metric $h_{\mu\nu}$ and the ghost-dilaton $d$ and the
corresponding sigma model fields, the string
metric $\tilde h_{\mu\nu}$ and dilaton $\Phi$. These relations are
needed to interpret
the tachyon vacuum solution and to discuss the possible
relation to the rolling solutions.

We begin by finding  the  precise linearized
relations between the string field dilaton and the sigma model
dilaton. The linearized relations  confirm
that the CSFT metric $h_{\mu\nu}$, which does not acquire an expectation
value in the tachyon vacuum, coincides with the string metric
of the sigma model, which does not change in the rolling solutions.
Moreover, the relation (\ref{dilaton relation}), together
with $h_{\mu\nu}=0$,
implies that our $d>0$ in the tachyon vacuum
 corresponds to $\Phi >0$, thus larger
string coupling.  This is also consistent with what we obtained
in the rolling solutions.

Our discussion of the linearized relations also allows us to
examine the various vertex operators associated with the various
dilaton fields used in the literature (section~3.2.).
In section 3.3 we examine the nonlinear relations between
the CSFT tachyon and dilaton and the effective field theory ones.
We work at zero momentum and up to quadratic order.  Finally,
in section 3.4, we present evidence that CSFT can describe
arbitrarily large dilaton deformations.

\subsection{Relating sigma model fields and string fields }

Consider first the effective action  (\ref{sigma_action}),
 suggested
by the conditions of conformal invariance of a sigma model
with gravity, dilaton and tachyon background fields.
If we set the tachyon to zero, this action reduces to the effective
action for massless fields, in the conventions of~\cite{Polchinski:1998rq}.
In this action $g_{\mu\nu}$ is  the string
metric, $\Phi$ is the diffeomorphism invariant
dilaton, and $T$, with potential
$V(T)=-\frac{2}{\alpha'} T^2+\cdots$, is the tachyon.
In order to compare  with the
string field action  we
expand the effective action in powers of small fluctuations using
\begin{equation}
g_{\mu\nu}=\eta_{\mu\nu}+\tilde h_{\mu\nu}\,,
\end{equation}
where we use a tilde in the fluctuation to distinguish it from the
metric fluctuation in the string field.  The result is
\begin{equation}
\label{sigma_limit}
\begin{split}
S_\sigma &=\frac{1}{2\kappa^2}\int d^D
x\,\Bigl(~ \frac{1}{4}\tilde h_{\mu\nu}\p^2 \tilde h^{\mu\nu}
-\frac{1}{4} \tilde h\p^2 \tilde h + \frac{1}{2}(\p^\nu \tilde
h_{\mu\nu})^2 + \frac{1}{2} \tilde h\p_\mu\p_\nu \tilde h^{\mu\nu}
\\[0.5ex]
&\qquad\qquad\qquad + 2\tilde h\,\p^2\Phi - 2\Phi\, \p_\mu\p_\nu \tilde
h^{\mu\nu} -4 \Phi\,\p^2 \Phi \\[0.6ex]
&\qquad\qquad\qquad - (\partial T)^2 + {4\over \alpha'} T^2
+\tilde h^{\mu\nu} \p_\mu
T\p_\nu T+ \bigl(\,{\tilde h\over 2} - 2\Phi\bigr)
(\partial T)^2 + \cdots  \Bigr)\,,
\end{split}
\end{equation}
where we have kept cubic terms coupling the dilaton and metric to
the tachyon. Such terms are needed to fix signs in the relations
between the fields in the sigma model and the string fields.

Let us now consider the string field action. The
string field needed to describe the tachyon, the metric fluctuations,
and the dilaton is
\begin{equation}
\label{scffc}
\begin{split}
|\Psi \ra &=\int \frac{d^D k}{(2\pi)^D} \Big( \,t(k) \, c_1 \bar c_1
-\frac{1}{2}
h_{\mu\nu}(k) \alpha_{-1}^\mu\bar \alpha_{-1} ^\nu c_1\bar c_1 +
d(k) (c_1 c_{-1}-\bar c_1\bar c_{-1})\\[0.5ex]
&\quad\quad\quad \quad+i \sqrt\frac{\alpha'}{2} B_\mu (k)
c_0^+(c_1\alpha_{-1}^\mu -\bar c_1\bar \alpha_{-1}^\mu) \Big)
|k\ra\,.
\end{split}
\end{equation}
Here $t(k)$ is the tachyon, $h_{\mu\nu}(k) = h_{\nu\mu}(k)$
is a metric fluctuation, $d(k)$ is the
ghost-dilaton, and $B_\mu(k)$ is an auxiliary field.
The sign and coefficient of $h_{\mu\nu}$ have been chosen
for future convenience. The linearized gauge
transformations of the component fields can be obtained from
$\delta |\Psi\ra =Q_B |\Lambda\ra$ with
\begin{equation}
|\Lambda\ra={i\over \sqrt{2\alpha'}} \, \epsilon_\mu
(c_1\alpha_{-1}^\mu- \bar c_1 \bar\alpha_{-1}^\mu) |p\ra\, .
\end{equation}
The resulting coordinate-space gauge
transformations are:
\begin{equation}
\label{gtcsf}
\delta h_{\mu\nu} =  \p_\nu
\epsilon_\mu+ \p_\mu \epsilon_\nu , \quad
\delta d =  -{1\over 2} \,\p\cdot \epsilon,
\quad
\delta B_\mu =- {1\over 2} \p^2\epsilon_\mu\,, \quad
\delta\, t = 0 \,   .
\end{equation}
We now calculate the quadratic part of the closed string field action,
finding
\begin{equation}
\label{messyaction}
\begin{split}
S^{(2)}&=-\frac{1}{\kappa^2\alpha'}~\la \Psi|c_0^- Q_B|\Psi\ra, \\
&=~\frac{1}{2\kappa^2}\int d^D x\,\Bigl(\,
\frac{1}{4}h_{\mu\nu}\p^2 h^{\mu\nu}- 2d\,\p^2 d\,-2 B_\mu (\p_\nu
h^{\mu\nu}+2\p^\mu d\,)-2B^2-
(\partial t)^2 + {4\over \alpha'} t^2 \Bigr),\\[0.5ex]
&=\frac{1}{2\kappa^2}\int d^D x\, \Bigl(
\,\frac{1}{4}h_{\mu\nu}\p^2 h^{\mu\nu}
+\frac{1}{2}(\p^\nu h_{\mu\nu})^2
-4 d\,\p^2 d
- 2d\,\p_\mu\p_\nu h^{\mu\nu} \,
- (\partial t)^2 + {4\over \alpha'} t^2
\Bigr)\,.
\end{split}
\end{equation}
In the last step we
eliminated the auxiliary field $B_\mu$ using its algebraic
equation of motion.

The gauge transformations (\ref{gtcsf}) imply that the
linear combination $d + {h\over 4}$ is  gauge
invariant. It follows that the sigma model dilaton must take the form
\begin{equation}
\label{dilaton relation 1} \lambda \, \Phi = d+ \frac{h}{4}\,,
\end{equation}
where $\lambda$ is a number to be determined.
Using (\ref{dilaton relation 1}) to eliminate the ghost-dilaton $d$
from the action (\ref{messyaction}) we find
\begin{equation}
\label{quadcsft}
\begin{split}
S^{(2)}&=\frac{1}{2\kappa^2}\int d^D x\, \Bigl(
~ \frac{1}{4}h_{\mu\nu}\p^2 h^{\mu\nu}
-\frac{1}{4} h\p^2 h
+  \frac{1}{2}(\p^\nu h_{\mu\nu})^2
+\frac{1}{2}  h\p_\mu\p_\nu h^{\mu\nu}\\[0.5ex]
&\qquad\qquad\qquad +\,2\lambda\, h\p^2\Phi -2\lambda\,
\Phi\,\p_\mu\p_\nu h^{\mu\nu} -4\lambda^2\, \Phi\p^2 \Phi
- (\partial t)^2 + {4\over \alpha'} t^2 \Bigr).
\end{split}
\end{equation}
We also use the string field theory to
calculate the on-shell coupling of $h_{\mu\nu}$ to two tachyons.
This coupling arises from the term
\begin{equation}
S^{(3)}=-\frac{1}{\alpha' \kappa^2}\la \,\mathcal{T}\,,\mathcal{H} \,,
\mathcal{T}\ra \,,
\end{equation}
where $\mathcal{T}$ and $\mathcal{H}$ denote the parts of the
string field (\ref{scffc}) that contain $t(k)$ and $h_{\mu\nu}(k)$,
respectively. We thus have
\begin{equation}
S^{(3)}
=\frac{1}{2\alpha' \kappa^2} \Bigl( \prod_{i=1}^3 \int
\frac{d^D k_i}{(2\pi)^D}\Bigr) \bigl\la c_1\bar c_1 e^{i k_1\cdot
X},c_1\bar c_1 \alpha_{-1}^\mu\bar\alpha_{-1}^\nu e^{i k_2\cdot
X}, c_1\bar c_1 e^{i k_3\cdot X} \bigr\ra \, t(k_1)t(k_3) h_{\mu\nu}(k_2)\,.
\end{equation}
The on-shell evaluation is readily carried out using $k^\mu h_{\mu\nu} (k)=0$.
We obtain
\begin{equation}
S^{(3)}=-\frac{1}{2\kappa^2} \int {d^D k_1\over (2\pi)^D}
{d^D k_3\over (2\pi)^D}
\, k_1^\mu k_3^\nu \,  t(k_1)t(k_3)
h_{\mu\nu}(-k_1-k_3)
= \frac{1}{2\kappa^2} \int d^D x\,h^{\mu\nu} \p_\mu t \p_\nu t\,.
\end{equation}
Combining this result with (\ref{quadcsft}) we obtain the
closed string field theory action
\begin{equation}
\label{quad+cubic_csft}
\begin{split}
S_{csft}&=\frac{1}{2\kappa^2}\int d^D x\, \Bigl(
~ \frac{1}{4}h_{\mu\nu}\p^2 h^{\mu\nu}
-\frac{1}{4} h\p^2 h
+  \frac{1}{2}(\p^\nu h_{\mu\nu})^2
+\frac{1}{2}  h\p_\mu\p_\nu h^{\mu\nu}\\[0.6ex]
&\qquad\qquad\qquad +\,2\lambda\, h\p^2\Phi -2\lambda\,
\Phi\p_\mu\p_\nu h^{\mu\nu} -4\lambda^2\, \Phi\p^2 \Phi  \\[0.6ex]
&\qquad\qquad\qquad
- (\partial t)^2 + {4\over \alpha'} t^2 +h^{\mu\nu} \p_\mu t \p_\nu t
+\ldots \Bigr).
\end{split}
\end{equation}

We are finally in a position to identify the sigma model action
(\ref{sigma_limit}) and the string field action (\ref{quad+cubic_csft}).
Comparing the quadratic terms in $\tilde h_{\mu\nu}$ and those in
$h_{\mu\nu}$ we see that $\tilde h_{\mu\nu} = \pm h_{\mu\nu}$.
We also note that $T= \pm t$.
The coupling $\tilde h^{\mu\nu} \p_\mu T \p_\nu T$ in (\ref{sigma_limit})
coincides with the corresponding coupling in (\ref{quad+cubic_csft})
if and only if
\begin{equation}
\tilde h_{\mu\nu}=h_{\mu\nu} \,.
\end{equation}
This simple equality justifies the multiplicative factor of $(-1/2)$
introduced for $h_{\mu\nu}$  in the string field (\ref{scffc}).  The
string field $h_{\mu\nu}$ so normalized is the fluctuation of the string
metric.
Comparing the couplings of metric and dilaton in both actions
we also conclude that $\lambda = +1$ and, therefore,
equation (\ref{dilaton relation 1}) gives
\begin{equation}
\label{dilaton relation}
  \Phi =
d+ \frac{h}{4}\,\,.
\end{equation}
This expresses the sigma model dilaton $\Phi$ in terms of the string
field metric trace and the ghost dilaton $d$.  It is important to
note that when we give a positive expectation value to $d$
(and no expectation value to $h$) we are increasing the value of $\Phi$
and therefore increasing the value of the string coupling.

\subsection {The many faces of the dilaton}

Equipped with the precise relations between string fields
and sigma-model fields we digress on the various
dilaton fields used in the literature.  Of particular
interest are the
corresponding vertex operators, which are determined by the
CFT states that multiply the component fields in the closed
string field.

We introduce the states
\begin{equation}
|\mathcal{O}^{\mu\nu}(p)\ra = -{1\over 4} ( \alpha_{-1}^\mu \bar\alpha_{-1}^\nu
+\alpha_{-1}^\nu \bar\alpha_{-1}^\mu) |p\ra \,, \quad
|\mathcal{O}^d(p)\ra = (c_1c_{-1}- \bar c_1 \bar c_{-1}) |p\ra\,.
\end{equation}
The corresponding vertex operators are
\begin{equation}
\mathcal{O}^{\mu\nu}(p) = {1\over 2\alpha'} ( \p X^\mu \bar\p X^\nu
+\p X^\nu \bar\p X^\mu)\, e^{ip X}, \quad
\mathcal{O}^d (p) = {1\over 2} (c \p^2 c- \bar c \bar\p^2 c) \, e^{ipX}\,.
\end{equation}
Working for fixed momentum, the string field (\ref{scffc}) restricted
to metric and dilaton
fluctuations~is
\begin{equation}
\label{mdfjhkj}
|\Psi\ra =  h_{\mu\nu}\,  |\mathcal{O}^{\mu\nu}\ra + d \, |\mathcal{O}^d\ra\,.
\end{equation}
This equation states that $\mathcal{O}^d$ is the vertex operator associated
with the ghost-dilaton field $d$. An excitation by this vertex operator does
not change the metric $h_{\mu\nu}$.
Our transformation to a gauge invariant dilaton gives
\begin{equation}
\label{firstred}
\Phi~ = d + {1\over 4} \, h\,, \quad
\tilde h_{\mu\nu} =  h_{\mu\nu} \,.
\end{equation}
Here $\tilde h_{\mu\nu}$ is the fluctuation of the
string metric.  Inverting these relations
\begin{equation}
\label{invfirstred}
d = \Phi - {1\over 4}\, \tilde h \,, \quad
  h_{\mu\nu} =  \tilde h_{\mu\nu} \,.
\end{equation}
Subtituting into the string field (\ref{mdfjhkj}) we obtain
\begin{equation}
\label{dflkkflk}
|\Psi\ra =  \tilde h_{\mu\nu} \Bigl(|\mathcal{O}^{\mu\nu}\ra - {1\over 4}\,
\eta^{\mu\nu} |\mathcal{O}^d\ra \Bigr)  + \Phi\,   |\mathcal{O}^{d}\ra\,.
\end{equation}
It is interesting to note that $\mathcal{O}^d$ is the vertex
operator associated with a variation of the gauge-invariant dilaton $\Phi$
and no variation of the string metric.  On the other hand,
$\mathcal{O}^{\mu\nu} - {1\over 4}\,
\eta^{\mu\nu} \,\mathcal{O}^d$  varies the
string metric and does not vary the gauge-invariant dilaton
(although it varies the ghost-dilaton).

Finally, we consider the formulation that uses the Einstein metric
$g_{\mu\nu}^E$
and the dilaton $\Phi$. The field redefinition is
\begin{equation}
g_{\mu\nu}^E = \exp (2\omega) \, g_{\mu\nu}\,,  \quad \hbox{with}
\quad \omega = -{2\over D-2}\, \Phi\,.
\end{equation}
Expanding in fluctuation fields we obtain
\begin{equation}
h_{\mu\nu}^E = \tilde h_{\mu\nu} - {4\over D-2} \, \eta_{\mu\nu} \, \Phi \,.
\end{equation}
Solving for $d$ and $h_{\mu}$ in terms of $\Phi$ and $h_{\mu\nu}^E$ we get
\begin{equation}
\label{invfirstredd}
d = -{2\over D-2} \,\Phi - {1\over 4} h^E \,, \quad
h_{\mu\nu} =  h_{\mu\nu}^E + {4\over D-2} \, \eta_{\mu\nu} \,
\Phi \,.
\end{equation}
Substituting into the string field (\ref{mdfjhkj}) we obtain
\begin{equation}
|\Psi\ra =   h_{\mu\nu}^E \Bigl( |\mathcal{O}^{\mu\nu}\ra - {1\over 4}\,
\eta^{\mu\nu} |\mathcal{O}^d\ra \Bigr)  +   {2\over D-2}\,\, \Phi\,
\Bigl( \, 2 \eta_{\mu\nu} |\mathcal{O}^{\mu\nu}\ra  -|\mathcal{O}^d\ra \Bigr)
\,.
\end{equation}
Interestingly, the vertex operator that varies the Einstein metric (without
variation of the dilaton) is the same as that for the
string metric~(see (\ref{dflkkflk})).
It is the dilaton operator that changes this time.  The vertex operator
\begin{equation}
\mathcal{D} =   2\eta_{\mu\nu}\mathcal{O}^{\mu\nu} - \mathcal{O}^d
=\Bigl( {2\over \alpha'}\, \p X \cdot \bar\p X
 - {1\over 2} (c \p^2 c- \bar c \bar\p^2 c) \,\Bigr) e^{ipX}\,,
\end{equation}
varies the dilaton without varying the Einstein metric.
This is the dilaton vertex operator used almost exclusively in
the early literature -- it is naturally associated with the Einstein
metric. The corresponding state $|\mathcal{D}(p) \ra$ has a particularly
nice property: it is annihilated by the  BRST operator when $p^2=0$. Indeed,
\begin{equation}
Q_B\,|\mathcal{D}(p)\ra = \frac{\alpha'}{2}p^2c_0^+ |\mathcal{D}(p)\ra \,.
\end{equation}
The dilaton $\mathcal{D}$ is in fact the unique linear
combination of the matter
and ghost dilatons that has this property. For other combinations,
terms linear in the momentum $p$ (such as $ (p\cdot \alpha_{-1})c_1\,\bar c_1 \bar
c_{-1} |p\rangle$),  survive.

\subsection{Relating the sigma model and string field dilaton and tachyon}

The closed string theory potential $V$, as read from
the effective action (\ref{sigma_action}) is
\begin{equation}
\label{Sigmamodelpot}
\kappa^2 V=e^{-2\Phi} \big( \, V (T) +\cdots
\big)\,, \quad \hbox{with} \quad V (T) = -T^2 + \cdots\,.
\end{equation}
Here $\Phi$ and $T$ are the zero momentum
dilaton and tachyon fields in the effective field theory.
The purpose of this section is to discuss the relation between
$\Phi$ and $T$ and the corresponding string fields
$d$ and $t$, both sets at zero-momentum. To do this we
must consider the effective potential for $d$ and $t$ calculated
in string field theory.  We only have the potential
itself. Collecting our previous results, we write
\begin{eqnarray}
\label{stringfieldform}
\kappa^2 V&=&-t^2+1.6018\, t^3-3.0172 \,t^4\nonumber\\[1.0ex]
&&+\,3.8721\, t^3 d
+(-0.8438\, t+1.3682\,t^2)\, d^2
-0.9528\, t\, d^3
-0.1056 \, d^4\, .
\end{eqnarray}
The contributions from massive fields affect quartic and higher
order terms.
In  our setup, the relevant terms arise when we eliminate
the level-four massive fields using their kinetic terms in
(\ref{gknfeh}) and their linear
couplings to $t^2$ in (\ref{fglekf4}), to
$td$ in (\ref{cgdt}), and to  $d^2$ in (\ref{cftcoudk}). We find
\begin{equation}
\Delta V = -{6241\over 186624}\, d^4 + {25329\over 16384}\, d^2\, t^2
- {1896129\over 4194304}\, t^4 \simeq
-0.0334\,  d^4 + 1.5460\, d^2 t^2 -  0.4521\, t^4\,.
\end{equation}
It follows that the effective potential for the tachyon
and the dilaton, calculated up to terms quartic in the fields and including
massive fields of level four only, is given by:
\begin{eqnarray}
\label{string field form}
\kappa^2 V_{eff}&=&-t^2+1.6018\, t^3-3.4693 \,t^4\nonumber\\[1.0ex]
&&+\,3.8721\, t^3 d
+(-0.8438\, t+2.9142\,t^2)\, d^2
-0.9528\, t\, d^3
-0.1390 \, d^4\, + \ldots\, .
\end{eqnarray}
The dots represent quintic and higher terms, which receive
contributions both from
elementary interactions and some integration of massive fields.
We write, more generically
\begin{eqnarray}
\label{stringxfield form}
\kappa^2 V_{eff} &=& - t^2+ a_{3,0} t^3+ a_{4,0} \,t^4\nonumber\\[1.0ex]
&&+a_{3,1}\, t^3 d
+(a_{1,2}\, t+a_{2,2}\,t^2)\, d^2
+ a_{1,3}\, t\, d^3
+ a_{0,4}\, d^4\,  + \ldots\, .
\end{eqnarray}
The values of the coefficients $a_{i,j}$ can be read comparing this
equation with (\ref{string field form}).

There are two facts about $V_{eff}$
 that make it
clear  it is not  in the form of
a ghost-dilaton exponential times a tachyon
potential. First,  it does not have a term
of the form $t^2 d$ that would arise from the tachyon mass
term and the expansion of the exponential.
Second, it contains a term linear in the tachyon; those terms
should be absent since the tachyon potential does not
have a linear term.
Nontrivial field redefinitions are necessary to relate
string fields and sigma model fields.

 To linearized order the
fields are the same, so we write relations of the form
 :
\begin{eqnarray}
\label{frsfsmf}
t&=&T+\alpha_1 T\Phi+\alpha_2 \Phi^2+\cdots, \nonumber\\[1.0ex]
d&=&\Phi+\beta_0 T^2 + \beta_1 T\Phi+\beta_2 \Phi^2 +\cdots\,,
\end{eqnarray}
where the dots indicate
terms of higher order in the sigma model fields.
We found no {\em need} for a $T^2$ term in the redefinition of tachyon
field, such a term would change the cubic and quartic self-couplings
of the tachyon in $V(T)$.
Since $d$ gives rise to  pure tachyon terms that are quadratic or higher,
only at quintic and higher order in $T$ will $V(T)$ differ from the
potential obtained by replacing $t\to T$ in the first line of
(\ref{string field form}).  We thus expect that after the field redefinition
(\ref{string field form}) becomes
\begin{equation}
\label{Sigmamodelpotx} \kappa^2 V=e^{-2\Phi} \big( -T^2  +1.6018\,
T^3-3.4693\,T^4 + \dots \big)\,,
\end{equation}
at least to quartic order in the fields.
We now plug the substitutions (\ref{frsfsmf})
into the potential (\ref{string field form}) and compare with
(\ref{Sigmamodelpotx}). A number of conditions emerge.
\begin{itemize}

\item In order to get the requisite $T^2\Phi$ term we need $\alpha_1 = -1$.

\item  In order to have a vanishing $T\Phi^2$ term $\alpha_2 = {1\over
2}a_{1,2}$ must be
half the coefficient of $td^2$ in (\ref{string field form}).

\item  Getting the correct
$T^3\Phi$ coupling then fixes $\beta_0 = (a_{3,0} - a_{3,1})/(2 a_{1,2})$.

\item  Getting the correct value of $T^2 \Phi^2$
fixes $\beta_1 = -(1+ {3\over 2} a_{3,0} a_{1,2} + a_{2,2})/ (2a_{1,2})$.
The vanishing of $T\Phi^3$
   fixes $\beta_2= - a_{1,3}/(2a_{1,2})$.  All coefficients in
    (\ref{frsfsmf}) are now fixed.

\item  The coefficient
        of $\Phi^4$, which should be zero, turns out to be  $(a_{0,4} + {1\over
4} a_{1,2}^2)\simeq
0.0389$,
which is small, but does not vanish.

\end{itemize}
Our inability to adjust the coefficient of $\Phi^4$ was to be expected.
The potential (\ref{string field form})
contains the terms $- t^2 +  a_{1,2} \, t d^2 + a_{0,4} d^4$
and, to this order, integrating out the tachyon gives an effective
dilaton quartic term of $(a_{0,4} + {1\over 4} a_{1,2}^2)$. With the
contribution
of the  massive fields beyond level four
this coefficient in the dilaton effective potential would vanish.
This is, in fact, the statement that was verified in~\cite{Yang:2005ep}.
It follows that we need not worry that the
quartic term in $\Phi$ do not vanish exactly.
Following the steps detailed before we
find
\begin{eqnarray}
\label{td to TD}
t&=&T- T\, \Phi-\, 0.4219 \, \Phi^2+\cdots, \nonumber\\[1.0ex]
d&=&\Phi  + 1.3453 \, T^2+  1.1180 \,T\, \Phi-\, 0.5646 \,
\Phi^2+\cdots.
\end{eqnarray}

\begin{figure}[!ht]
\leavevmode
\begin{center}
\epsfysize=6.0cm
\epsfbox{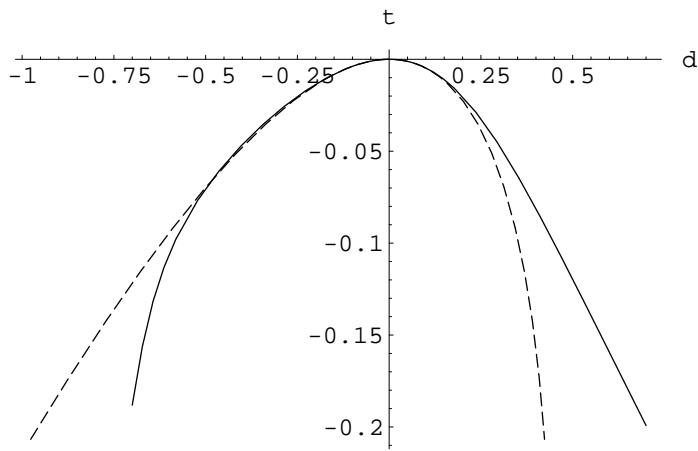}
\end{center}
\caption{\small The solid line is the dilaton marginal
direction  defined by the set of points $(d, t(d))$ where $t(d)$ is the
expectation
value of $t$ obtained solving the tachyon equation of motion for
the given $d$. The dashed line represents the direction along
the sigma model dilaton $\Phi$ (thus $T=0$). It is obtained by setting
$T=0$ in equation (\ref{td to TD}).
The two lines agree well even reasonably far from the origin.}
\label{fig marginal directions}
\end{figure}


In string field theory the dilaton deformation is represented
in the $(d,t)$ plane
by the curve $(d, t(d))$, where $t(d)$ is the expectation
value of the tachyon when the dilaton is set equal to $d$.
This curve, calculated using the action (\ref{string field form}),
  is shown as a  solid line in Figure~\ref{fig marginal directions}.
On the other hand, it is clear  that $\Phi$ (with $T=0$) defines the
marginal direction in the effective field theory. Setting $T=0$ in
(\ref{td to TD}) we find the pair $(d(\Phi), t(\Phi))$, which must
be a parameterization of the flat direction in terms of $\Phi$.
This curve is shown as a dashed line in Figure~\ref{fig marginal
directions}. It is a good consistency check that these two
curves agree well with each other over a significant fraction of
the plot.

\subsection{Dilaton deformations}

In Ref.~\cite{Yang:2005ep} we computed the effective dilaton potential
that arises when we integrate out the tachyon from a potential
that includes only quadratic and cubic terms.  We found that the domain
of definition of this potential is the full real $d$ line. This
happens because the (marginal) branch $t(d)$ that gives the
expectation value of $t$ for a given value of $d$ is well defined
for all values of $d$.  In this section we extend this computation
by including higher level fields and higher order interactions.
As we will demonstrate, it appears plausible  that the domain
of definition for  the effective dilaton potential remains $d\in (-\infty,
\infty)$.

The marginal branch is easily identified for small values
of the dilaton: as the dilaton expectation value goes to zero
all expectation values go to zero.  For large enough values of
the dilaton the marginal branch may cease to exist, or it may
meet another solution branch.   If so, we obtain limits on the
value of $d$.  Since the dilaton effective potential is supposed
to be flat in the limit of high level, we propose the following
criterion.
If we encounter a limit value of $d$, this value is deemed
reliable only if the dilaton potential at this point is  not
very large.  A large value for the potential indicates that the calculation
is not reliable because the same terms that are needed to make the potential
small could well affect the limit value.  In open string field theory
a reliable limit value was obtained for the Wilson line parameter:
at the limit point the potential energy density was a relatively small fraction
of the D-brane energy density.
The purely cubic potential for $t$ gives a critical point
with $\kappa^2 V \sim - 0.05774$.
We define $\mathcal{R}(d) \equiv\frac{|\kappa^2 V(d)|}{0.05774}$, where $V(d)$
is the
effective dilaton potential. A critical value of $d$ for which $\mathcal{R} >
1$
will be considered unreliable.

We start with cubic potentials and then include the elementary
quartic interactions level by level. With cubic potentials, the
effective dilaton potential
is invariant under $d\to -d$. With
$\mathbb{V}^{(3)}_4$ dilaton deformations  can be arbitrarily
large~\cite{Yang:2005ep}.  We then find

\begin{itemize}
\item The dilaton potential derived
from  $\mathbb{V}^{(3)}_8$ is defined
for  $|d|\leq 624$. This is plausible since, at this level,
the equations of motion for the level-four fields are linear.

\item  The dilaton potential derived
from  $\mathbb{V}^{(3)}_{12}$ is defined
for $|d|\leq 1.71$. Since  $\mathcal{R}(\pm 1.71)=42.4$,
there is no reliable limit value.

\item The dilaton potential derived from  $\mathbb{V}^{(4)}_{0}$
is defined
for $|d|\leq 4.67$, where $\mathcal{R}(\pm
4.67)=49.5$. The large value of $\mathcal{R}$ indicates
that there is no evidence of a limit value.

\item The dilaton potential derived from $\mathbb{V}^{(4)}_{2}$ is
not  invariant under $d\to -d$. We find a
range  $d \in (-\infty\,, 3.124)$ . Although
$\mathcal{R}(3.124)=0.387$, the potential has a maximum with
$\mathcal{R}=3.325$ at $d=1.92$. This fact makes the limit point
$d= 3.124$ unreliable.

\item The dilaton potential derived from $\mathbb{V}^{(4)}_{4}$, the highest
level potential we have computed fully, is regular
for $d\in (-2.643,  \,6.415)$. Since $\mathcal{R}(6.415)=1502.4$
and $\mathcal{R}(-2.643)=89.2$,  there is no
branch cut in the reliable region.
\end{itemize}

%

We have also computed  the higher level
quartic interactions $t d^3$ and
$d^4$.  We have checked  that
$\mathbb{V}^{(4)}_{4}$, supplemented by those interactions
does not lead to branch cuts in the potential for the dilaton.
This result, however, is not conclusive.  Additional
interactions must be included at level six (the level of $t d^3$)
and at level eight (the level of $d^4$).

\begin{figure}[!ht]
\leavevmode
\begin{center}
\epsfysize=6.5cm
\epsfbox{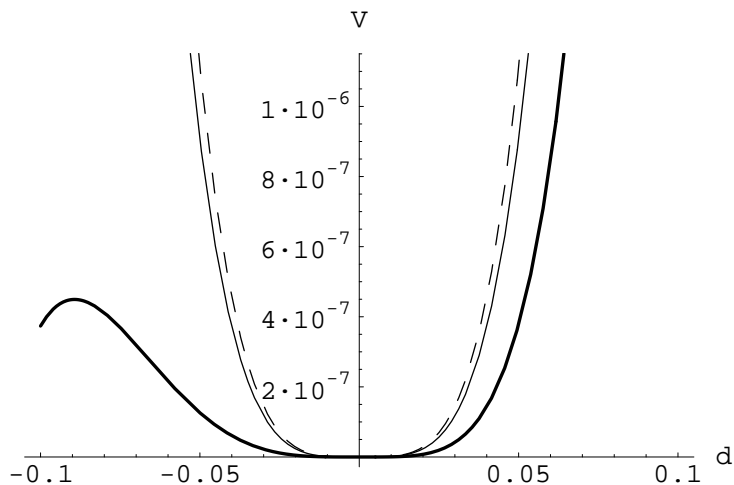}
\end{center}
\caption{\small Dilaton effective potential. The dashed line
arises from $\mathbb{V}^{(3)}_{4}$, the solid line arises from
$\mathbb{V}^{(3)}_{8}$, and the thick line arises from
$\mathbb{V}^{(4)}_{8}$.} \label{dilaton potentials}
\end{figure}

We tested in~\cite{Yang:2005ep} that cubic and quartic
interactions combine to give a vanishing quartic term in the
dilaton effective potential. We can ask if the potential for the
dilaton becomes flatter as the level of the calculation is
increased. We find that it roughly does, but the major changes in
the potential are due to the elementary quartic term in the
dilaton. For the cubic vertex, the interactions of the type $d^2
M$, with $M$ massive give rise to terms quartic on the dilaton.
Other cubic couplings that do not involve the dilaton typically
induce $d^6$ (and higher order) terms, which play a secondary role
in flattening the potential if the quartic terms have not
cancelled completely. Therefore, the potentials that arise from
$\mathbb{V}^{(3)}_{8}$, $\mathbb{V}^{(3)}_{10}$ and
$\mathbb{V}^{(3)}_{12}$ (without the contribution from level six
massive fields) have no obvious difference. The potentials
obtained at various levels are shown in Figure~\ref{dilaton
potentials}. The dashed line arises from $\mathbb{V}^{(3)}_{4}$,
the solid line arises from $\mathbb{V}^{(3)}_{8}$, and the thick
line arises from $\mathbb{V}^{(4)}_{8}$.

\sectiono{Conclusions}

In this paper we have presented some calculations that
suggest the existence of a tachyon vacuum for the bulk
closed string tachyon of bosonic string theory.  We have
discussed the physical interpretation using the effective
field theory both to suggest the value of the action
density at the critical point (zero!) and to obtain
rolling solutions~\cite{HZR} that seem consistent with the interpretation
of the tachyon vacuum as a state in which there are no
closed string states.

The numerical evidence presented is still far from conclusive.
A critical point seems to exist and appears to be robust, but it
is not all that clear what will happen when the accuracy of
the computation is increased.  If the action density at the
critical point goes to zero it may indeed define a
new and nontrivial tachyon vacuum.
Conceivably, however, the critical point could approach the
perturbative vacuum, in which case there would be no evidence
for a new vacuum.  Alternatively, if the action density at the
critical point remains finite, we would have no interpretation
for the result.  

\begin{figure}[!ht]
\leavevmode
\begin{center}
\epsfysize=9.1cm
\epsfbox{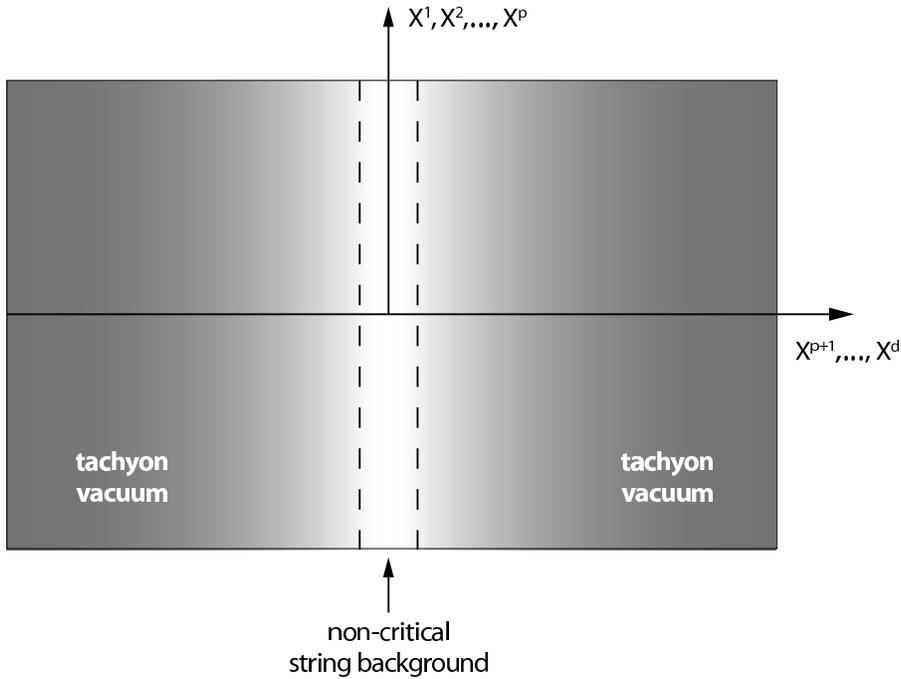}
\end{center}
\caption{\small A non-critical $(p+1)$-dimensional
string theory would
correspont to a solitonic solution of critical string
theory in which, far away from the reduced space, the fields
approach the values of the closed string tachyon vacuum.}
\label{ncrit_conj}
\end{figure}

Let us consider some additional indirect arguments that support the
existence of a closed string tachyon vacuum.  The first one arises
from the existence of sub-critical bosonic string theories. The
evidence in string theory is that most string theories are related
by compactifications and/or deformations.
It seems very likely that non-critical string theories are
also related to critical string theory.  It should then
be possible to obtain a non-critical string theory as a solution
of critical string theory. Certainly the view that $D=2$ bosonic
string theory is a ground state of the bosonic string has been
held as likely~\cite{2dconj}.  In non-critical string theory the number of
space dimensions is reduced (at the expense of a linear dilaton
background).  The analogy with lower-dimensional D-branes in open
string theory seems apt: the branes are solitons of the open string
field theory tachyon in which far away from the branes the tachyon
sits at the vacuum.  It seems plausible that non-critical string
theories are solitonic solutions of the {\em closed} string theory
tachyon.  As sketched in Figure~\ref{ncrit_conj}, far away
along the coordinates transverse to the non-critical
world-volume, the background would approach the
closed string tachyon vacuum.  The universality of the tachyon vacuum
would imply that a noncritical string theory could be further
reduced using the same background configuration used to reduce the original
critical theory.

In fact,  in the $p$-adic open/closed string theory lump solutions
of the closed string sector appear to describe spacetimes of lower
dimensionality, as explained by
Moeller and Schnabl~\cite{Moeller:2003gg}.
Indeed, far away from the lump the open string tachyon must be at its
vacuum and therefore there are no D-brane solutions with more space dimensions
than
those of the lump. Away from the lump the closed string
tachyon is at its vacuum, and no linearized solutions of the equations
of motion exist.

A suggestive argument for zero action at the tachyon vacuum follows
from the sigma model approach.  As discussed by
Tseytlin~\cite{Tseytlin:2000mt}, it seems likely that
the closed string effective action for the spacetime background fields
may be written in terms of the partition function $Z$ of the two-dimensional
sigma model as well as derivatives thereof (this does work for open
strings~\cite{Witten:1992cr}). The conventional  coupling of the
world-sheet area to the tachyon $T$ results
in a partition function and an effective action  with a prefactor of $e^{-T}$.
Thus one expects a tachyon potential of the form $e^{-T} g(T)$ where $g$ is a
polynomial that begins with a negative quadratic term\footnote{In
\cite{Tseytlin:2000mt},
a tachyon potential of the form $-T^2 e^{-T}$ is considered. Complications
in fixing the kinetic terms made it unclear if $T=\infty$
was a point in the configuration space (see the discussion below eqn.~(4.13))
of \cite{Tseytlin:2000mt}. For additional comments on the possible
form of the tachyon potential, see Andreev~\cite{Andreev:2003hn}.}. In
this case, for a tachyon vacuum at  $T\to\infty$ the action goes to zero.

The computations and the discussion presented in this paper
have led to a set of testable conjectures concerning  the vacuum of
the bulk closed string tachyon of bosonic string theory.  It seems
likely that additional computations, using both string field theory,
effective field theory, and conformal field theory  will help test these ideas
in the near future.

\bigskip
\noindent
{\bf Acknowledgements} We are grateful to M.~Headrick
and A.~Sen for many
instructive discussions.  We would also
like to acknowledge useful conversations with K.~Hashimoto,
H. Liu,  N. Moeller, Y.~Okawa, M. Schnabl, and  A. Tseytlin.

\appendix

\sectiono{Quartic Computations} \label{a1}

\subsection{The setup}
We normalize correlators using $ \langle 0| c_{-1}\bar c_{-1} c_0^- c_0^+ c_1
\bar c_1|0\rangle=1$
with $c_0^\pm = {1\over 2} (c_0 \pm \bar c_0)$.
All states  in this paper have zero momentum.
For convenience, all spacetime coordinates have been compactified and the
volume of spacetime is equal to one.
To use  results from open
string field theory, we note that
\begin{equation}
\label{faccorr}
\langle  c(z_1)c(z_2)c(z_3) \,   \bar c (\bar w_1)\bar c (\bar
w_2)\bar c (\bar w_3)
\rangle = -2 \langle  c(z_1)c(z_2)c(z_3) \rangle_o \,\cdot\,\langle
\bar c (\bar w_1)\bar c (\bar w_2)\bar c (\bar w_3) \rangle_o\,,
\end{equation}
since open string field theory uses
$\langle  c(z_1)c(z_2)c(z_3) \rangle_o = (z_1-z_2)(z_1-z_3) (z_2 - z_3)$.
Then:
\begin{equation}
\langle c_1\bar c_1,c_1\bar c_1,c_1\bar
c_1\rangle= 2\cdot \langle c_1,c_1,c_1\rangle_o\
\cdot \langle \bar c_1,\bar c_1,\bar
c_1\rangle_o = 2\cdot \mathcal{R}^3 \cdot \mathcal{R}^3 = 2 \mathcal{R}^6\,,
\end{equation}
where $\mathcal{R} \equiv 1/
\rho = 3\sqrt{3}/ 4 \simeq 1.2990\,,$
and $\rho$ is the  mapping radius of the disks in the
three-string vertex.

To construct four-string amplitudes we use
antighost insertions~\cite{Zwiebach:1992ie,Belopolsky:1994sk}
\begin{equation}
\label{exptwoforms}
\mathcal{B} = \sum_{I=1}^4 \sum_{m=-1}^\infty
(B^{I}_m\, b^J_m + \overline{ \Ci^I_m }\, {\bar b}^I_m) \,, \quad
\mathcal{B}^\star = \sum_{I=1}^4 \sum_{m=-1}^\infty
(\Ci^I_m\, b^I_m + \overline{B^I_m} \, {\bar b}^I_m)\,,
\end{equation}
where $\mathcal{B}^\star$ is the $\star$-conjugate of
$\mathcal{B}$. The multilinear function in string field theory is
\begin{equation}
\label{amplitude}
\{ \Psi_1, \Psi_2, \Psi_3, \Psi_4\} \equiv
{1\over \pi} \int_{\mathcal{V}_{0,4}} \hskip-3pt dx\wedge dy  \,
\langle \Sigma | \,\mathcal{B} \, \mathcal{B}^\star
\,|\Psi_1\rangle |\Psi_2\rangle|\Psi_3\rangle|\Psi_4\rangle\,.
\end{equation}
The first, second, third, and fourth states are inserted at
$0,1,\xi = x+ iy,$ and $\infty$, respectively.
Operationally, the fourth state is inserted
at $t=0$ with  $z = 1/t$, where $z$
is the global uniformizer.  For further details and
explanations the reader should consult~\cite{Yang:2005ep}.
We record that
\begin{equation}
\label{minusonecoeff}
\begin{split}
B_{-1}^{J}& =\delta_{3J}/\rho_3,\quad
\Ci_{-1}^{J}=0 \,, \\
B_1^{I}~&=\rho_I\p \beta_I+ {\textstyle{1\over 2}}
\rho_3\varepsilon_3\delta_{I3}\,,\quad
\Ci_1^{I}=\rho_I\bar\p \beta_I\,,
\\[0.8ex]
B_2^I~&={\textstyle{1\over 6}}\rho_I^2 \p ( 2\beta_I^2 - \varepsilon_I)
+ \rho_I^2 (-4 \delta_I-2 \varepsilon_I \beta_I+ 8 \beta_I^3)\delta_{3I},
\quad  \Ci_2^I={\textstyle{1\over 6}}
\rho_I^2 \bar\p (2\beta_I^2- \varepsilon_I)\,.
\end{split}
\end{equation}
Here $\bar\partial \equiv {\partial \over
\partial \bar \xi}$ and $\partial \equiv {\partial
\over \partial \xi} $. Since our string fields are annihilated
both by $b_0$ and $\bar b_0$, the coefficients $B^I_0$ and $\Ci^I_0$
are not needed.
   Taking note of the vanishing coefficients,
we see that for states in the Siegel gauge the antighost factor $\mathcal{B}$
is given by
\begin{equation}
\label{Bgeneral}
\mathcal{B} = B_{-1}^{3} b_{-1}^{(3)} +
\sum_{I=1}^4
(B^{I}_1\, b^J_1 + \overline{ \Ci^I_1 }\, {\bar b}^I_1)  +
\sum_{I=1}^4
(B^{I}_2\, b^J_2 + \overline{ \Ci^I_2 }\, {\bar b}^I_2) + \ldots  \,.
\end{equation}

The Strebel quadratic differential on the surfaces determines:
\begin{equation}
\label{betavalues}
\beta_1={a\over 2\xi} -{1\over \xi} -1 \,,\quad
\beta_2=\frac{a-2\xi}{2(1-\xi)}\,,\quad
\beta_3=\frac{a-2}{2\xi(\xi-1)}\,,\quad
\beta_4= {a\over 2} -1 -\xi \,.
\end{equation}
Here $a(\xi, \bar \xi)$ is a function that determines
the quadratic differential completely.  We also have
\begin{equation}
\label{epsvalues}
\begin{split}
\varepsilon_1&= 2+{1\over \xi} (a-2) + {1\over \xi^2}
\Bigl( 2 + a - {5\over 8} a^2\Bigr)\,, \\[0.3ex]
\varepsilon_2&= {-5a^2 + 16\, \xi(\xi -3) + 8a \,(\xi + 3)\over
8\, (\xi-1)^2}\,,
\\[0.3ex]
\varepsilon_3&= {16+ 8a -5a^2 + 24(a-2)\xi \over 8\,\xi^2 \,(\xi-1)^2}\,,
\\[0.3ex]
\varepsilon_4&= 2+ a - {5\over  8} \, a^2 - 2 \xi + a\, \xi + 2 \xi^2 \,.
\end{split}
\end{equation}

The function $a(\xi)$ is known
numerically to high accuracy for $\xi\in \mathcal{A}$, where
$\mathcal{A}$ is a specific subspace of $\mathcal{V}_{0,4}$
described in detail in Figures 3 and 6 of ref.~\cite{Moeller:2004yy}. The
full space $\mathcal{V}_{0,4}$ is obtained by acting on
$\mathcal{A}$ with the transformations generated by
$\xi\to 1-\xi$ and $\xi\to 1/\xi$, together with complex
conjugation $\xi\to
\bar\xi$. In fact $\mathcal{V}_{0,4}$
contains twelve copies of $\mathcal{A}$.  Let $f(\mathcal{A})$
denote the region obtained by mapping each point $\xi\in
\mathcal{A}$ to $f(\xi)$. Then $\mathcal{V}_{0,4}$ is composed of the six
regions
\begin{equation}
\label{various_regions}
\mathcal{A} \,, ~~ {1\over \mathcal{A}}\,,~~ 1-\mathcal{A}\,,
   ~~{1\over 1-\mathcal{A}} \,, ~~1- {1\over \mathcal{A}}\,,~~
{\mathcal{A}\over 1-\mathcal{A}} \,,
\end{equation}
together with their complex conjugates.  The values of $a$ in these
regions follow from the values of $a$ on $\mathcal{A}$ via the relations
\begin{equation}
a\,(1-\xi)= 4-a(\xi)\,,\quad
a\Bigl(\,{1\over\xi}\,\Bigr)= {a(\xi)\over \xi}\,,\quad
a(\,\bar\xi\,)=\overline{a(\xi)} \,.
\end{equation}

\medskip
For states of the form
$|M_i\rangle  = \mathcal{O}_i  c_1 \bar c_1 |0\rangle$,
where $\mathcal{O}_i$ is  built with
matter oscillator, one finds
\begin{equation}
\{ M_1, M_2, M_3, M_4\} = -{2\over \pi }\int_{\mathcal{V}_{0,4}}
{dx\wedge dy\over
(\rho_1\rho_2\rho_3\rho_4)^2}  \,\langle\langle
\mathcal{O}_1\mathcal{O}_2\mathcal{O}_3
\mathcal{O}_4\rangle\rangle_\xi\,.
\end{equation}
Here
$
\langle\langle \mathcal{O}_1\mathcal{O}_2\mathcal{O}_3
\mathcal{O}_4\rangle\rangle_\xi \equiv \langle h_1\circ \mathcal{O}_1
~ h_2\circ
\mathcal{O}_2 ~h_3\circ \mathcal{O}_3 ~h_4\circ \mathcal{O}_4
\rangle_{\Sigma_\xi},
$
where the right-hand side is a matter correlator computed after
the local operators $\mathcal{O}_i$ have been mapped to the uniformizer.

\subsection{Couplings of dilatons and tachyons}

\underbar{Elementary contribution to $t^3 d$}.
We insert the dilaton on the moving puncture to make the
integration identical over each of the 12 regions of the moduli
space.
Since all the states inserted on the fixed punctures have ghost
oscillators $c_1\bar c_1$, the antighost factor $\mathcal{B}\,
\mathcal{B}^\star$ is only supported on the moving puncture:
\begin{eqnarray}
\mathcal{B}\, \mathcal{B}^\star (c_1 c_{-1} -\bar c_1\bar
c_{-1})^{(3)}|0\rangle=-(B_{-1}^{3}
\Ci_1^{3}+\overline{B_{-1}^{3}}\, \overline{ \Ci_1^{3}})|0\rangle
=-(\bar\p \beta_3+\p\bar\beta_3)|0\rangle.
\end{eqnarray}
There are no matter operators, thus the correlator just involves
the ghosts:
\begin{eqnarray}
\langle \Sigma_P|\mathcal{B}\, \mathcal{B}^\star |T\rangle
|T\rangle |D\rangle |T\rangle&=&-(\bar\p \beta_3+\p\bar
\beta_3)\langle(c_1\bar c_1)^{(1)} (c_1\bar c_1)^{(2)}  (c_1\bar
c_1)^{(4)} \rangle\nonumber\\[1.0ex]
&=& -(\bar\p \beta_3+\p\bar \beta_3) {2\over
(\rho_1\rho_2\rho_4)^2}\,.
\end{eqnarray}
Using (\ref{amplitude}), the amplitude is:
\begin{equation}
\{T^3D\} =-\frac{24}{\pi}\int_{\mathcal{A}} dx dy\,
(\bar\p \beta_3+\p\bar \beta_3) \frac{1}{(\rho_1 \rho_2 \rho_4)^2}
=\,23.2323\, .
\end{equation}
The contribution to the potential is
$\kappa^2 V=\frac{4}{4!}\{T^3D\}\,t^3 d =3.8721\,   t^3 d\, .$

\medskip
\noindent\underbar{Elementary contribution to $t^2 d^2$}. We
insert the dilatons at
$z_2=1$ and $z_3=\xi$. The amplitude
to be integrated is identical to the ghost part of the amplitude
for the quartic interaction $a^2 d^2$, as given
in~\cite{Yang:2005ep},
equation (4.9):
\begin{equation}
\label{ghpart2dil}
\langle\Sigma|\mathcal{B}\mathcal{B}^\star |T\rangle |D\rangle |D\rangle
|T\rangle = {2\over (\rho_1\rho_4)^2} \Bigl(  \bar\p \beta_2
\p(\bar\xi \bar \beta_3) -
\p\beta_2 \bar\p ( \bar \xi \bar \beta_3) + *\hbox{-conj}\Bigr)\,.
\end{equation}
The four-point amplitude is then
\begin{eqnarray}
\{T^2D^2\}=\frac{4}{\pi}\int_{\mathcal{V}_{0,4}} {dx  dy\over
(\rho_1\rho_4)^2} ~ \hbox{Re}\Big(\bar\p \beta_2 \p(\bar\xi \bar \beta_3) -
\p\beta_2 \bar\p ( \bar \xi \bar \beta_3)\Big)\,.
\end{eqnarray}
Since we have the same states on punctures one and four, and these
punctures are exchanged by the transformation $z\to 1/z$, the
integral over $\mathcal{A}$ gives the same contribution as the
integral over $1/\mathcal{A}$. The conjugation
properties of the amplitude also imply that
$\overline{\mathcal{A}}$ contributes the same as $\mathcal{A}$.
Consequently, the four regions $\mathcal{A},\, 1/\mathcal{A},
\,\overline{\mathcal{A}},$ and $1/\overline{\mathcal{A}}$ all give
the same contribution.  To get the full amplitude we must multiply
the contributions of $\mathcal{A}$, of $1-\mathcal{A}$, and
$1-1/\mathcal{A}$ by four:
\begin{eqnarray}
\{T^2D^2\}=4\cdot \frac{4}{\pi}\,\Bigl[\int_{\mathcal{A}}
+\int_{1-\mathcal{A}}+
\int_{1-1/\mathcal{A}}\Bigr]{dx  dy\over
(\rho_1\rho_4)^2} ~ \hbox{Re}\Big(\bar\p \beta_2 \p(\bar\xi \bar \beta_3) -
\p\beta_2 \bar\p ( \bar \xi \bar \beta_3)\Big).
\end{eqnarray}
The transformation laws given in Appendix B of~\cite{Yang:2005ep}
allow one to rewrite
     the second
and third integrals as integrals over $\mathcal{A}$, where they can be easily
evaluated. We find
\begin{eqnarray}
\{T^2D^2\}=4\cdot (-0.2410 +0.4031+1.2065)
=5.4726.
\end{eqnarray}
The contribution to potential is
$\kappa^2 V=\frac{6}{4!}\{T^2D^2\}\,t^2 d^2 =1.3682 \, t^2 d^2.$

\medskip
\noindent
\underbar{Elementary contribution to $t d^3$}.
The  tachyon field is inserted at $z_3=\xi$.  We then have
\begin{eqnarray}
&& \mathcal{B}\, \mathcal{B}^\star (c_1 \bar c_1)^{(3)}D^{(1)}
D^{(2)}
D^{(4)}|0\rangle \nonumber\\[1.0ex]
&=&\hskip-5pt\Big\{B_{-1}^{3}b_{-1}^{(3)}+\sum_{J\not= 3} \big( B_1^{J}
b_1^{(J)} +\overline{\Ci_1^{J}}\, \bar b_1^{(J)}\big)\Big\}
\Big\{\overline{B_{-1}^{3}}\,\bar
b_{-1}^{(3)}+\sum_{J\not= 3} \big( \overline{B_1^{J}}\,\bar
b_1^{(J)} +\Ci_1^{J} b_1^{(J)}\big)\Big\}(c_1 \bar
c_1)^{(3)}D^{(1)} D^{(2)}
D^{(4)}|0\rangle \nonumber\\[1.0ex]
&=&\hskip-9pt\sum_{I\not= J\not= K\not=3} \Bigl(
{\textstyle{1\over 2}} B_{-1}^{3}
\Ci_1^{I}  D^{(J)} \, D^{(K)}\, c_1^{(I)} \bar c_1^{(3)} +
B_1^{I} \Ci_1^{J} (c_1\bar
c_1)^{(3)} c_1^{(I)} c_1^{(J)} (\bar
c_{-1} \bar c_1)^{(K)}\Big)|0\rangle  -\star\hbox{-conj}\,.
\end{eqnarray}
Therefore, the correlator  $\mathcal{C}_{td^3} =
\langle \Sigma| \mathcal{B}\, \mathcal{B}^\star
TD^3|0\rangle$ is:
\begin{equation*}
\mathcal{C}_{td^3}
=\hskip-8pt\sum_{I\not= J\not= K\not=3}\Bigl\langle
 -B_{-1}^{3}
\Ci_1^{I}   (\bar c_{-1} \bar c_1)^{(J)} (c_{-1} c_1)^{(K)}\,
c_1^{(I)} \bar c_1^{(3)}\, + \,
B_1^{I} \Ci_1^{J}   (\bar
c_{-1} \bar c_1)^{(K)}c_1^{(I)} c_1^{(J)}(c_1\bar
c_1)^{(3)}\Bigr\rangle + *\hbox{-conj}\,.
\end{equation*}
Factorizing into holomorphic and antiholomorphic parts we get
\begin{equation}
\mathcal{C}_{td^3}
=2\hskip-8pt\sum_{I\not= J\not= K\not=3}\Bigl(
 B_{-1}^{3}
\Ci_1^{I} B_{KI} (B_{J3})^*  - \,
B_1^{I} \Ci_1^{J} D_{IJ} (B_{K3})^* \Bigr) + *\hbox{-conj},
\end{equation}
where $B_{IJ} \equiv \langle (c_{-1}c_1)^{(I)}, c_1^{(J)} \rangle$
was introduced and evaluated in~\cite{Yang:2005ep}, eqns. (4.18), (4.20),
and (4.21). Additionally,
\begin{equation}
D_{IJ} \equiv \langle c_1^{(I)}, c_1^{(J)}, c_1^{(3)} \rangle =
{z_{IJ} z_{I3}z_{J3}\over \rho_I \rho_J \rho_3},
\qquad  D_{I4} = - D_{4I} = {z_{I3} \over \rho_I \rho_3\rho_4} \,,
\quad I, J  \not=4\,.
\end{equation}
The full amplitude is
\begin{equation}
\{TD^3\}= \frac{12}{\pi}\int_{\mathcal{A}} dx dy\,\mathcal{C}_{td^3}
=\,-5.7168 \,.
\end{equation}
The contribution to the potential is
$\kappa^2 V=\frac{4}{4!}\{TD^3\}\, t d^3=-0.9528\, t d^3.$

\subsection{Couplings of tachyon to massive fields}

\noindent  In all cases the massive field will be inserted
on the moving puncture $z_3=\xi$.

\noindent
\underbar{Elementary contribution to $t^3 f_1$}.
With $F_1 \equiv  c_{-1} \bar c_{-1}$ inserted at $z_3=\xi$
we find:
\begin{equation}
\mathcal{B}\, \mathcal{B}^\star (c_{-1} \bar
c_{-1})^{(3)}|0\rangle =(\Ci_1^{3}
\overline{\Ci_1^{3}}-B_1^{3}\, \overline{B_1^{3}} )|0\rangle.
\end{equation}
\begin{equation}
\{T^3F_1\}=\frac{12}{\pi}\int_{\mathcal{A}} dx dy\,
\frac{2}{(\rho_1\rho_2\rho_4)^2}(\Ci_1^{3}
\overline{\Ci_1^{3}}-B_1^{3}\, \overline{B_1^{3}} )
=-2.6261\, .
\end{equation}
The contribution to the potential is:
$\kappa^2 V= \frac{4}{4!}\{T^3F_1\}t^3 f_1 =-0.4377 \,t^3 f_1.$

\medskip
\noindent
\underbar{Elementary contribution to $t^3 f_2$}.
With $F_2 \equiv c_{1} \bar c_{1}L_{-2}\bar L_{-2}$
at $z_3=\xi$, the
ghost part is that of the four-tachyon amplitude
(eqn. (3.34) of~\cite{Yang:2005ep}).
With $w=0$ corresponding to $z=z_3$, and $S(z,w)$ denoting
the Schwarzian derivative, the holomorphic matter correlator
is:
\begin{equation}
\begin{split}
\langle L_{-2}^{(3)}\rangle &=\langle
T^{(3)}(w=0)\rangle
=\rho_3^2 \langle  T(z_3)\rangle
+ {26\over 12} S(z,w)
=\frac{13}{6}\rho_3^2(2\beta_3^2-\varepsilon_3)\,.
\end{split}
\end{equation}
Therefore, the amplitude is
\begin{equation}
\{T^3F_2\} =-\frac{24}{\pi}\int_{\mathcal{A}}
\,\frac{dx dy}{(\rho_1\rho_2\rho_3\rho_4)^2}
\Big|\frac{13}{6}\rho_3^2(2\beta_3^2-\varepsilon_3)\Big|^2
=-\,337.571.
\end{equation}
The contribution to the potential is
$\kappa^2 V= \frac{4}{4!}\{T^3F_2\} t^3 f_2=-56.262\,  t^3 f_2.$

\medskip
\noindent \underbar{Elementary contribution to $t^3 f_3$}. With
$L_{-2} c_1\bar c_{-1}$ inserted at $z_3=\xi$ we find
\begin{equation}
\mathcal{B}\, \mathcal{B}^\star (c_1 \bar
c_{-1})^{(3)}|0\rangle=-B_{-1}^{3}
\overline{B_1^{3}}|0\rangle  \,.
\end{equation}
\begin{equation}
\mathcal{C}_{t^3f_3} \equiv
  \langle \Sigma | \mathcal{B}\, \mathcal{B}^\star TT
(c_1 \bar c_{-1})^{(3)} T |0\rangle \cdot \langle L_{-2}^{(3)} \rangle
=- \frac{2B_{-1}^{3}
\overline{B_1^{3}}}{(\rho_1\rho_2\rho_4)^2} \cdot
\frac{13}{6}\rho_3^2(2\beta_3^2-\varepsilon_3)\,.
\end{equation}
With $F_3 \equiv L_{-2} c_1\bar
c_{-1} + c_{-1} \bar L_{-2} \bar c_1$, the  string amplitude
relevant to $t^3 f_3$ is:
\begin{equation}
\{T^3\, F_3\}
=\frac{12}{\pi}\int_{\mathcal{A}} dx dy\,
(\mathcal{C}_{t^3f_3} + \mathcal{C}_{t^3f_3}^*)
=78.1432\,.
\end{equation}
The contribution to
the potential is: $\kappa^2 V=\frac{4}{4!}\{T^3F_3\}\, t^3f_3=13.024\, t^3f_3.$

\medskip
\noindent \underbar{Elementary contribution to $t^3 g_1$}. With
$b_{-2} c_1\, \bar c_{-2}\bar c_1$ at $z_3=\xi$, one finds
\begin{equation}
   \mathcal{B}\, \mathcal{B}^\star(b_{-2} c_1 \bar c_{-2} \bar
c_1)^{(3)}|0\rangle
=\overline {\Ci_2^{3}} \, \, \overline{ B_{-1}^{3}}  (c_1 b_{-2}
)^{(3)}  |0\rangle.
\end{equation}
The state $ c_1b_{-2}|0\rangle$ is created by the non-primary
ghost current $j (z) = cb (z)$ by acting on the vacuum. For the ghost current
\begin{equation}
j(w) = j(z) {dz\over dw} -{3\over 2} {z''\over z'}
\quad \to \quad j(w=0)= \rho_3 ( j(z_3) - 3 \beta_3) \,.
\end{equation}
We thus have the correlator:
\begin{equation}
\begin{split}
\mathcal{C}_{t^3g_1} ~\equiv & ~~\langle \Sigma|
\mathcal{B}\, \mathcal{B}^\star T T (b_{-2} c_1\, \bar c_{-2}\bar c_1)^{(3)}
 T |0\rangle  \\[1.0ex]
=& ~~\overline {\Ci_2^{3}} \, \,
\overline{ B_{-1}^{3}} \,\frac{1}{(\rho_1\rho_2\rho_4)^2}
\Bigl\langle c\bar c (0)  \, c\bar c (1) \rho_3 ( j(z_3) - 3 \beta_3)
c\bar c (t=0) \Bigr\rangle \\[1.0ex]
=& ~~\overline {\Ci_2^{3}} \, \,
\overline{ B_{-1}^{3}} \,\frac{\rho_3}{(\rho_1\rho_2\rho_4)^2}
\cdot
2 \Big(\frac{1}{\xi}+\frac{1}{\xi-1}-3\beta_3\Big)\,.
\end{split}
\end{equation}
With $G_1 \equiv b_{-2} c_1\, \bar c_{-2}\bar c_1-
c_{-2}c_1\bar b_{-2}\bar c_1$, the amplitude relevant
for the $t^3 g_1$ coupling is
\begin{equation}
\{T^3\, G_1\}
=\frac{12}{\pi}\int_{\mathcal{A}} dx dy\,
(\mathcal{C}_{t^3g_1} + \mathcal{C}_{t^3g_1}^*)
=1.6350\,.
\end{equation}
The contribution to the potential is $
\kappa^2 V=\frac{4}{4!}\{T^3G_1\}\, t^3g_1 =0.2725 \, t^3g_1$.

\end{document}